\documentclass{jfm}
\usepackage{subcaption}
\usepackage{graphicx}
\usepackage{epstopdf,epsfig}

\newcommand{\ang}[1]{\ensuremath{\left\langle {#1} \right\rangle}}
\newcommand{\kmin}{k_\mathrm{min}}
\newcommand{\kmax}{k_\mathrm{max}}
\newcommand{\tstar}{\mathcal{T}}
\newcommand{\ens}{\ang{\omega^2}}
\newcommand{\ke}{\ang{E_k}}
\newcommand{\es}{\ang{E_\gamma}}
\newcommand{\dint}{\zeta}

\title{Droplet-turbulence interactions and quasi-equilibrium dynamics in turbulent emulsions}
\author{Siddhartha Mukherjee\aff{1},
		Arman Safdari\aff{2},
		Orest Shardt\aff{2},
		Sa\v{s}a Kenjere\v{s}\aff{1},
	Harry E.A. Van den Akker\aff{1,2}\corresp{\email{Harry.VanDenAkker@tudelft.nl}}
		}

\affiliation{\aff{1}Section of Transport Phenomena, Department of Chemical Engineering, Delft University of Technology,
Van der Maasweg 9, 2629HZ, Delft, Netherlands
\aff{2} Bernal Institute, School of Engineering, Faculty of Science and Engineering, University of Limerick, Limerick, Ireland}

\begin{document}

\maketitle

\begin{abstract}
We perform direct numerical simulations (DNSs) of emulsions in homogeneous, isotropic turbulence using a pseudopotential lattice-Boltzmann (PP-LB) method. Improving on previous literature by minimizing droplet dissolution and spurious currents, we show that the PP-LB technique is capable of long, stable simulations in certain parameter regions. Varying the dispersed phase volume fraction $\phi$, we demonstrate that droplet breakup extracts kinetic energy from the larger scales while injecting energy into the smaller scales, increasingly with higher $\phi$, with the Hinze scale dividing the two effects. Droplet size ($d$) distribution was found to follow the $d^{-10/3}$ scaling \citep{deane2002scale}. We show the need to maintain a separation of the turbulence forcing scale and domain size to prevent the formation of large connected regions of the dispersed phase. For the first time, we show that turbulent emulsions evolve into a quasi-equilibrium cycle of alternating coalescence and breakup dominated processes. Studying the system in its state-space comprising kinetic energy $E_k$, enstrophy $\omega^2$ and the droplet number density $N_d$, we find that their dynamics resemble limit-cycles with a time delay. Extreme values in the evolution of $E_k$ manifest in the evolution of $\omega^2$ and $N_d$ with a delay of $\sim0.3\tstar$ and $\sim0.9\tstar$ respectively (with $\tstar$ the large eddy timescale). Lastly, we also show that flow topology of turbulence in an emulsion is significantly more different than single-phase turbulence than previously thought. In particular, vortex compression and axial straining mechanisms become dominant in the droplet phase, a consequence of the elastic behaviour of droplet interfaces.
\end{abstract}


\section{Introduction}
An emulsion consists of a dense suspension of droplets of one fluid (the dispersed phase) suspended in another fluid (the continuous phase), and is formed due to turbulent mixing of these two immiscible fluids. Emulsions are found (both desirably and undesirably) in a wide range of industries. For instance, in food processing, diverse products depend on the stability and texture of emulsions \citep{mcclements2015food}. In biotechnology, emulsions can serve as miniature laboratories where living cells can be compartmentalized into individual droplets \citep{griffiths2006miniaturising}. They are also known to cause various losses in crude oil production \citep{kokal2005crude}, or to the contrary, enable enhanced oil recovery \citep{banat1995biosurfactants}. The formation of an emulsion requires shearing of droplets which can occur both in laminar and turbulent flows, although the latter may be a more common occurrence. Turbulent emulsions can be said to form a particular class of droplet laden turbulent flows where there is close interplay between turbulence and the dynamics of the dispersed fluid. An accurate description of these systems hence involves the dynamics of deforming interfaces, while allowing for coalescence and breakup of droplets, resolution of a range of length and time scales of turbulent flow and the possible presence of surface active agents (surfactants) that can alter the interfacial dynamics.

The primary effect of turbulence on droplets in these flows is to cause fragmentation, where an initially large connected volume of the dispersed phase is broken into smaller droplets. Under sustained turbulence, there is a supposed equilibrium between coalescence and breakup which leads to a droplet spectrum around a theoretical maximum stable diameter, known as the Hinze scale \citep{hinze1955fundamentals}. This droplet distribution is also known to follow a $d^{-10/3}$ slope \citep{deane2002scale}, where $d$ is the droplet diameter. The dispersed phase influences the continuous phase turbulence by drawing turbulent kinetic energy (TKE) from the flow, which partially goes into the difference between the surface energy of parent and daughter droplets, while the rest is stored in the deformation of interfaces. This reduces the effective turbulent kinetic energy (TKE), which has consequences on the turbulence cascade and spectrum, noticeably at scales comparable to droplet sizes. Coalescing droplets in turn set finer flow structures into motion, where interfacial tension releases the energy stored in droplet deformations back as TKE into the flow \citep{dodd2016interaction} at scales smaller than the droplet sizes.

\subsubsection*{Literature review}
In this paper, we are interested in simulating an emulsion under sustained turbulence. Simulations are key here, as experiments as yet are incapable of revealing the underlying dynamical processes - limited by emulsions being optically opaque, while interfacial dynamics is inherently three-dimensional, whereby experiments give mostly statistical or phenomenological results. There have been only a few numerical studies devoted to turbulent emulsion dynamics, some of which have been detailed in the recent review by \cite{elghobashi2019dns} on DNS simulations of turbulent flows laden with droplets or bubbles. We refer interested readers to it for a general overview, while we shall discuss the current state of simulating turbulent emulsions, highlighting those aspects that we intend to address in our work.

In one of the first studies, \cite{derksen2007multi} simulated a turbulent liquid-liquid dispersion using a free-energy based lattice-Boltzmann (LB) method. They modeled a fluid packet as it passes by the impeller in a stirred vessel, hence experiencing a burst of turbulence, before entering a quiescent zone. They show evolution of the droplet distribution in the dispersion under first constant and then decaying turbulence, also reporting the modification to the kinetic energy spectra at a crossover scale.

\cite{perlekar2012droplet} simulated droplet breakup in homogeneous, isotropic turbulence using a pseudopotential (PP) LB method, showing that the distribution of droplet diameters has a finite width around the Hinze scale. Since Hinze's criterion does not account for droplet coalescence or coagulation, deviation from it was found at higher volume fractions. Further, droplet breakup was attributed to peaks in the local energy dissipation rate. The study reported on the method being originally incapable of attaining steady state simulations due to droplet dissolution, which was remedied by a mass correction scheme to artificially re-inflate droplets which helped maintain a steady volume fraction \citep{biferale2011lattice}. Later, \cite{perlekar2014spinodal} simulated turbulent spinodal decomposition to show coarsening arrest in a symmetric binary fluid mixture (which is similar to an emulsion, although the morphology is distinctly different). The presence of turbulence was shown to inhibit the coarsening dynamics at droplet sizes larger than the Hinze scale.
 
\cite{skartlien2013droplet} simulated a surfactant laden emulsion under weak turbulence ($Re_\lambda \leq 20$) using a free-energy LB method, and reproduced a $d^{-10/3}$ droplet distribution as predicted by \cite{deane2002scale} for air bubbles above the Hinze scale. They did not find any influence of the surfactant in altering the coalescence rates in the considered range of surfactant activity and turbulence intensity. Also using a free-energy LB method \cite{komrakova2015numerical} simulated turbulent liquid-liquid dispersions at varying volume fractions, focusing on the resolution of droplets with respect to the Kolmogorov scale. They found that droplet dissolution was a significant issue, which made it impossible to obtain a steady state droplet distribution at low phase fractions, while at higher phase fractions ($\phi>0.2$), despite breakup, most droplets coalesce to form a single connected region with multiple smaller satellite droplets. Increasing the resolution of the Kolmogorov scale remedied droplet dissolution to some extent, and a log-normal droplet distribution was shown from transient simulations, as has been experimentally found for turbulent liquid-liquid dispersions \citep{pacek1998sauter,lovick2005drop}. The multiphase energy spectra could not be reproduced due to spurious currents which caused unphysical energy gain at high wavenumbers, whose magnitude was found to be close to the turbulent velocity scale $u^\prime$. 

In their detailed study on droplet-turbulence interaction, \cite{dodd2016interaction} simulated a large number of initially spherical droplets ($\phi=0.05$) in decaying homogeneous, isotropic turbulence using a mass conserving volume-of-fluid method. They considered a wide range of density and viscosity ratios between the droplet and carrier fluid, and showed an enhanced rate of energy dissipation for increasing droplet Weber number ($W\!e$). Introducing the TKE equations, they show that breakup and coalescence act as source and sink terms of TKE. \cite{roccon2017viscosity} studied the influence of viscosity on breakup and coalescence in a swarm of droplets ($\phi=0.18$) in wall bounded turbulent flow using a coupled Cahn-Hillard Navier-Stokes solver. They report a slight drag reduction in the flow due to the presence of droplets, and show that a higher interfacial tension or droplet viscosity favours coalescence, and the number of droplets rapidly decreases to $1-10\%$ of its initial value. At low viscosity, where breakup dominates, around $50\%$ of the droplets remain separated and their sizes follow Hinze's $\left\langle D\right\rangle \propto W\!e^{-3/5}$ criterion. 

Recently, using a mass conserving level-set method, \cite{shao2018direct} studied interface-turbulence interactions in droplet breakup simulations. They showed that vortical structures tend to align with large scale interfaces before breakup. They also show that there is a slight increase in axial straining and vortex compression in the flow topology in the presence of droplets, in comparison to single-phase turbulence.

\subsubsection*{Our study}
In this study, we resolve several of the issues faced in previous work, and report new findings from direct numerical simulations of turbulent emulsions. We use the PP-LB method for a multicomponent fluid system without phase change to simulate the formation of a dispersion starting from a single large drop in the center of a periodic box. PP-LB is well suited for the simulation of multiphase flows comprising deformable droplets due to the spontaneous formation of interfaces emerging from simplified interparticle repulsion forces \citep{shan1993lattice,shan1994simulation,shan1995multicomponent}. The method has been used before for simulating droplets in turbulence \citep{perlekar2012droplet,perlekar2014spinodal,albernaz2017droplet}, an in general has been applied many more times for non-turbulent flows with droplets. It allows for coalescence and breakup to occur naturally, all without the need for interface tracking or models for film drainage. However, it comes with a caveat that due to interfaces being diffuse, coalescence is favourable when interfaces overlap. This makes the resolution of interface width relative to droplet sizes, i.e. the Cahn number, an important criterion \citep{shardt2013simulations}. The diffuse interface also leads to dissolution of small droplets as has been noted before \citep{perlekar2012droplet,komrakova2015numerical}. We show that droplet dissolution can be limited to a minor effect in certain parameter ranges, and that a mass correction scheme as used in \cite{biferale2011lattice,perlekar2012droplet} is not requisite for simulating droplets in turbulence.

Additionally, multiphase LB simulations suffer from spurious currents ($u^{\mathrm{sp}}$) which are velocities arising from anisotropy in the discretization of inter-particle forces. Although $u^{\mathrm{sp}}$ in pseudopotential LB have been shown to be much lower than in conventional finite volume techniques like the volume-of-fluid method \citep{kamali2013simulating,mukherjee2018simulating}, in the free-energy LB method they were found strong enough to dominate the multiphase kinetic energy spectra at high wavenumbers \citep{komrakova2015numerical}. Further, in LB, the characteristic fluid velocity (here the large scale velocity $\mathcal{U}$) should be kept smaller than the lattice speed of sound $c_s$, such that the flow Mach number $M\!a = \mathcal{U}/c_s$ is low (where traditionally $M\!a<0.3$ is considered incompressible) and hence the flow being simulated obeys the incompressible Navier-Stokes equations. Hence, the velocities should scale as $c_s > \mathcal{U} \gg u^{\mathrm{sp}}$, which we maintain in our work.

We simulate a dispersion in a periodic box, employing a forcing scheme to generate homogeneous, isotropic turbulence. Here we particularly study the influence of varying the dispersed phase volume fraction ($\phi$) and turbulence intensity ($Re_\lambda$) on the properties of the dispersion formed. We show the influence of the dispersed phase on the multiphase kinetic energy spectra which has not been systematically presented before, or was not possible due to the limitations of the numerical method \citep{komrakova2015numerical}. We show that $\phi$, $Re_\lambda$ and the interfacial tension $\gamma$ together determine the dispersion morphology, and that droplets of a particular characteristic length can be generated by varying these parameters. Investigating local flow topology, we show that the effect of the dispersed phase is significant and more pronounced than previously stated \citep{shao2018direct}, with a sharp increase in vortex compression and axial straining in the droplet regions. We also present, for the first time, an analysis of the equilibrium dynamics of a droplet laden isotropic turbulent flow, showing that the system evolution in its state-space is akin to a limit-cycle with alternating dominance of coalescence and breakup as the system oscillates between different dispersion morphologies.

\subsubsection*{Length scales}
Through this study we highlight a few considerations that have not been discussed in previous work and are crucial to simulating droplets in turbulence. First is numerically resolving to a sufficient degree the several length scales that govern different aspects of these simulations. The main length scale is the typical droplet diameter, which can be taken to be the Hinze scale and is given as \citep{hinze1955fundamentals}
\begin{equation}
d_{\mathrm{max}} = 0.725(\rho^c/\gamma)^{-3/5}\epsilon^{-2/5}
\end{equation}
where $\rho^c$, $\gamma$ and $\epsilon$ are the carrier fluid density, interfacial tension and rate of energy dissipation, respectively, while it is now accepted that the local variations in $\epsilon$ (intermittency) set a local Hinze scale, and an entire spectrum of droplets centered around $d_{\mathrm{max}}$ tends to arise. A closely associated length scale is the interface width $\dint$, which in physical systems can be of the order of nanometers for micron to millimeter size droplets. However, as a limitation of our simulation technique (and every other diffuse interface method), the interface width extends over a few computational grid cells. The ratio between $\dint$ and the droplet diameter $d$ is termed the Cahn number $Ch = \dint/d$ \citep{komrakova2015effects}, and extreme values of $Ch$ are undesirable. Hence the relative separation between $d$ and $\dint$ needs to be considered. 

Next, the two length scales characterizing turbulence are the energy injection scale $\mathcal{L}$ which is determined by the forcing scheme, and the smallest (or Kolmogorov) scale $\eta$ which is determined by the viscosity $\nu$ and the dissipation rate $\epsilon$. A wide separation between $\mathcal{L}$ and $\eta$ means a higher Reynolds number $Re$, which can be expressed as $Re \approx \left( \mathcal{L}/\eta\right)^{4/3}$. A final length scale of importance in simulations is the size of the simulation domain, which along one spatial direction can be considered to be $N_x$, and this is generally chosen to be close to $\mathcal{L}$. As droplets will break up due to extension under turbulent stresses, the domain size $N_x$ should be sufficiently larger than the maximum droplet elongation before breakup to yield meaningful results (particularly for simulations on periodic domains, where large droplets would begin to interact with images of themselves). Here a particular caveat is also the simplistic description of highly deformed droplets, where an equivalent droplet diameter $d=(6V/\pi)^{1/3}$ gives the impression of $N_x \gg d$, whereas in the form of long, slender filaments, droplets can linearly extend across the entire domain. This can give rise to elongated droplets that remain connected due to periodicity, and this is more prone to occur at high volume fractions under weak turbulence, as for instance can be seen in \cite{skartlien2013droplet}.

Comparing these length scales, the required spatial separation between them for simulating droplets in the inertial range, at least from a stance of reasoning, would follow as 
\begin{equation}
N_x \gg \mathcal{L} \gg d \gg \eta \gg \dint
\end{equation}
while $N_x>\mathcal{L}$ may also be sufficient, and most studies currently are limited to $N_x \approx \mathcal{L}$. Also, $d$ can vary over a range of values, extending upto $d \sim \eta$ if the Kolmogorov scale is over-resolved. Upon conceding to limitations of modeling, current simulations can at best reproduce
\begin{equation}
N_x > \mathcal{L} \gg d \gg \eta \approx \dint
\end{equation}
We try to maintain such a separation of scales in the study, except that we have $\dint > \eta$. Lastly, having $\eta>d$ would mean sub-Kolmogorov droplets. These droplets can also deform and breakup due to the action of viscous stresses instead of inertial stresses \citep{elghobashi2019dns}. 

We begin with a description of the numerical method in section \ref{sec:NumericalMethod}, followed by a brief validation of the turbulence forcing scheme. We then present results from turbulent emulsions in section \ref{sec:turbulentEmulsions}, for varying volume fraction in section \ref{sec:volFracVary} and varying turbulence intensity in section \ref{sec:turbIntensity}. Section \ref{sec:domainSize} discusses the importance of sufficient resolution of the largest scales and section \ref{sec:forcingWavenumber} shows the influence of the turbulence forcing wavenumber on the dispersion morphology. Finally, in section \ref{sec:dynamics} we discuss some general results regarding emulsion dynamics, with the quasi-equilibrium limit-cycle presented in section \ref{sec:limitCycle}, droplet-vorticity alignment in section \ref{sec:vorticityAlign} and influence of droplets on local flow topology in section \ref{sec:topology}, after which we end with the conclusions. 

\section{Numerical Method}\label{sec:NumericalMethod}
\subsection{Lattice Boltzmann Method}
Each component $\sigma \in \left\lbrace\alpha,\beta\right\rbrace$ obeys the standard LBGK equation with a single relaxation time which can be written as \citep{kruger2017lattice}
\begin{eqnarray}
f_i^\sigma\left(\mathbf{x} +\mathbf{c}_i\Delta t, t + \Delta t\right) & = & f_i^\sigma\left(\mathbf{x},t\right) - \frac{f_i^\sigma\left(\mathbf{x},t\right) - f_i^{\mathrm{eq},\sigma}\left(\mathbf{x},t\right)}{\tau^\sigma}\Delta t 
\end{eqnarray}
where $f_i^\sigma$ is the distribution function of component $\sigma$ along the discrete velocity direction $\mathbf{c}_i$. Here $\tau^\sigma$ is the lattice relaxation time towards local equilibrium which relates to the macroscopic component viscosity $\nu^\sigma = c_s^2(\tau^\sigma - 1/2)$ where $c_s = 1/\sqrt{3}$ is the lattice speed of sound (the mixture viscosity is a more complex expression when the components have different $\tau$). The equilibrium distribution $f_i^{\mathrm{eq},\sigma}$ is given by the local Maxwellian as
\begin{equation}
f_i^{\mathrm{eq},\sigma} = w_i\rho \left( 1 + \frac{\mathbf{u}^{\mathrm{eq}}\cdot \mathbf{c}_i}{c_s^2} + \frac{\left(\mathbf{u}^{\mathrm{eq}}\cdot \mathbf{c}_i\right)^2}{2c_s^4} - \frac{\mathbf{u}^{\mathrm{eq}}\cdot \mathbf{u}^{\mathrm{eq}}}{2c_s^2}\right)
\end{equation}
where $w_i$ are the LB weights in each direction $i$, and $\mathbf{u}^{\mathrm{eq}}$ is the equilibrium velocity which is given as
\begin{equation}
\mathbf{u}^{\mathrm{eq}} = \mathbf{u}^\prime + \frac{\tau^\sigma \mathbf{F}^\sigma}{\rho^\sigma}
\end{equation}
Here $\mathbf{F}^\sigma$ incorporates all the forces (here the inter-component interactions and the turbulence forcing), into the \textit{common} fluid velocity $\mathbf{u}^\prime$ between the two components which is given as
\begin{equation}
\mathbf{u}^\prime = \displaystyle\frac{\sum_\sigma \displaystyle\frac{\rho^\sigma\mathbf{u}^\sigma}{\tau^\sigma}}{\sum_\sigma\displaystyle\frac{\rho^\sigma}{\tau^\sigma}}
\end{equation}
where $\mathbf{u}^\sigma$ is the \textit{bare} component velocity. This is calculated in its usual form
\begin{equation}
\mathbf{u}^\sigma = \frac{1}{\rho^\sigma}\sum_i f_i^\sigma \mathbf{c}_i 
\end{equation}
For details see \cite{succi2001lattice,kruger2017lattice}. The inter-component interaction force, $\mathbf{F}^{\mathrm{SC}}$, is modeled using the method of \cite{shan1995multicomponent}, which can be written as
\begin{equation}
\mathbf{F}^{\mathrm{SC},\sigma}\left( \mathbf{x}\right) = - G_{\sigma \overline{\sigma}}\psi^\sigma \left(\mathbf{x}\right) \sum_{\sigma\neq\overline{\sigma}} \psi^{\overline{\sigma}} \left(\mathbf{x} + \mathbf{c}_i\Delta t\right)\mathbf{c}_i w_i\Delta t
\end{equation}
where $\psi^\sigma$ is the pseudopotential function for component $\sigma$ and in this study we have chosen $\psi^\sigma = \rho^\sigma$ (while other definitions are possible). This force between the components is kept to be repulsive, hence the interaction strength parameter $G_{\sigma\overline{\sigma}}$ should have a positive value. It should be noted that the fluids remain partially miscible, and essentially the final composition consists of $\alpha-$rich and $\beta-$rich regions, while a small amount of one component remains dissolved in the other. A higher magnitude of $G_{\sigma\overline{\sigma}}$ results in lower solubility and gives rise to a higher interfacial tension. The equation of state for this multicomponent system is \citep{kruger2017lattice}
\begin{equation}
p = c_s^2 \sum_\sigma \rho^{\sigma} + \frac{c_s^2 \Delta t^2}{2}\sum_{\sigma,\overline{\sigma}}G_{\sigma\overline{\sigma}}\psi^{\sigma}\psi^{\overline{\sigma}}
\end{equation}
Lastly, the interfacial tension $\gamma$ can be calculated using the Laplace law $\Delta p = 2\gamma/r$, where $\Delta p$ is the pressure difference across the interface of a spherical droplet.

The simulations here have been performed on a $D3Q19$ lattice, i.e. a three-dimensional lattice with a set of $19$ discrete velocity directions. Further, the lattice spacing $\Delta x$ and time step $\Delta t$ are both set equal to $1$, and consequently all quantities are expressed in dimensionless lattice units [lu].

\subsection{Turbulence Forcing}
To generate and sustain turbulence in the fluid, a constant source of energy is required, which is constantly being dissipated by viscosity at the smallest scales (i.e. the Kolmogorov scales). This is done by setting the largest scales of flow into motion, and if the fluid viscosity is low enough, these large structures become unstable and give rise to successively smaller scales. One of the ways to achieve this numerically is by employing a low wavenumber spectral forcing, as given by \cite{alvelius1999random}, while alternative techniques could also be used \citep{eswaran1988examination,rosales2005linear}. This forcing was also implemented by \cite{ten2004fully} in LB to simulate the response of clouds of spherical solid particles to homogeneous isotropic turbulence. A very similar form of the forcing is used by \cite{perlekar2012droplet}, which is constructed directly in real space but could be made to have a similar effective spectral form as \citep{ten2004fully,ten2006application}, albeit with less control over output parameters, as we do in this study. The forcing is divergence free by construction and can be written as
\begin{eqnarray}
F_x^\sigma &= \sum_{k={k_a}}^{{k_b}} \displaystyle\frac{\rho^\sigma}{\rho^{\mathrm{tot}}} A(k) \left[ \sin(2\pi k y + \phi_y(k)) + \sin(2\pi k z + \phi_z(k)) \right] \nonumber \\
F_y^\sigma &= \sum_{k={k_a}}^{{k_b}} \displaystyle\frac{\rho^\sigma}{\rho^{\mathrm{tot}}} A(k) \left[ \sin(2\pi k x + \phi_x(k)) + \sin(2\pi k z + \phi_z(k)) \right] \nonumber \\
F_z^\sigma &= \sum_{k={k_a}}^{{k_b}} \displaystyle\frac{\rho^\sigma}{\rho^{\mathrm{tot}}} A(k) \left[ \sin(2\pi k x + \phi_x(k)) + \sin(2\pi k y + \phi_y(k)) \right]
\label{eq:forcing}
\end{eqnarray}
Here $\rho^{\mathrm{tot}} = \sum_\sigma \rho^\sigma$ is the total density considering both components and each $\phi_i(k)$ is a unique random phase. Alternatively, $\phi_i(k)$ can be evolved as a stochastic process, as done in \cite{perlekar2012droplet}, but in our approach $\phi_i(k)$ (and hence the forcing) varies as white noise in time. This ensures that the force is not related to any timescale of turbulent motion, and is a choice also made in \cite{ten2006application}. The force is distributed over a small range of wavenumbers $k_a \leq k \leq k_b$, while the contribution of each of these wavenumbers is determined by $A(k)$ which centers the Gaussian around $k_f$ in Fourier space, given as
\begin{equation}
A(k) = A\exp\left( -\frac{\left(k -k_f\right)^2}{c}\right)
\end{equation}
where $k_f$ is the central forcing wavenumber, $c$ is a width over which to distribute the force amplitude and is set to $c=1.25$, and $A$ is a forcing magnitude. This method ensures that there is a dominant central wavenumber $k_f$ (which can also be a fraction) in the forcing scheme, while neighbouring wavenumbers also contain some energy, which makes the scheme more stable \citep{ten2006application}. Lastly, the total power input to the fluid can be written as the sum of two terms as follows
\begin{equation}
P = P_1 + P_2 = \frac{1}{2}\overline{f_k f_k}\Delta t + \overline{u_k f_k}
\end{equation}
where the two terms are the force-force and force-velocity correlations respectively, and $u_k,f_k$ refer to the volumetric velocity and force fields. The force-velocity correlation, $P_2$, should be $0$ to avoid an uncontrolled growth of energy in the fluid \citep{alvelius1999random}, and it is achieved by varying the force term at each time step. This is computationally expensive, hence some studies \citep{ten2004fully,ten2006application} vary the force by choosing randomly from a pre-computed set of force fields at each time step. This was found to introduce a non-zero contribution from the $P_2$ term, where the steady state kinetic energy was roughly $10$ times larger than with a unique random force at each time step - hence in this study we adhere to the latter approach.

The largest scale in the system is given by the domain size $N_x$, which sets the minimum wavenumber $\kmin = 2\pi/N_x$. All other wavenumbers are integer multiples of $\kmin$, with the maximum wavenumber being $\kmax = \kmin N_x/2 = \pi$. The smallest scale of turbulence (Kolmogorov scale) is calculated as $\eta \sim \left( \nu^3/\epsilon\right)^{1/4}$ where $\nu$ and $\epsilon$ are the kinematic viscosity and energy dissipation rate respectively. The criterion for a resolved DNS simulation is that $\kmax\eta > 1$ \citep{moin1998direct}, and the Kolmogorov scale should obey $\eta>0.318$ [lu] \citep{ten2006application}. We shall mention the forcing wavenumber $k_f$ and the wavenumber bounds as multiples of $\kmin$ in this study. For a central forcing wavenumber $k_f$, the associated large scale length then becomes 
\begin{equation}
\mathcal{L} \sim \frac{2\pi}{k_f\kmin} = \frac{N_x}{k_f}
\end{equation}
Further, the Taylor microscale is calculated as 
\begin{equation}
\lambda = \left( \frac{15\nu {u^\prime}^2}{\epsilon}\right)^{1/2}
\label{eq:TaylorLambda}
\end{equation}
where $u^\prime$ is the root mean square velocity along one direction, and $u^\prime_x = u^\prime_y = u^\prime_z$ in isotropic turbulence. The rate of energy dissipation $\ang{\epsilon}$ can be found in two ways, as $\epsilon \approx \nu\ang{\omega^2} \approx \sum_k 2\nu k^2 E(k)/N_x^3$ where $\ang{\omega^2}$ is the average enstrophy and $E(k)$ is the kinetic energy spectrum. Using $\lambda$, the Taylor Reynolds number is calculated as
\begin{equation}
Re_\lambda = \frac{u^\prime \lambda}{\nu}
\label{eq:TaylorRe}
\end{equation}
Lastly, the Kolmogorov timescale is given as
\begin{equation}
\tau_k = \left(\frac{\epsilon}{\nu}\right)^{-1/2}
\end{equation}
For eddies in the inertial range with a size $l$, the velocity $u(l)$ and timescale $\tau(l)$ are determined uniquely by $\epsilon$ and $l$ alone as $u(l) = (\epsilon l)^{1/3} \sim \mathcal{U} (l/\mathcal{L})^{1/3}$ and $\tau(l) = (l^2/\epsilon)^{1/3} \sim \mathcal{T}(l/\mathcal{L})^{2/3}$, where $\mathcal{L}$, $\mathcal{T}$ and $\mathcal{U}$ are the characteristic length, time and velocity of the largest eddies (with $\mathcal{T} = \mathcal{L}/\mathcal{U}$). We consider $\mathcal{U} \approx {\ang{E_k}}^{1/2}$ as the largest eddies contain most of the kinetic energy, and generally $u^\prime < \mathcal{U}$. The characteristic velocity at a particular length scale can also be found from the kinetic energy spectrum as $u(l) \approx \sqrt{E(k_l)}$ where $k_l = 2\pi/l$.

\section{Single-phase turbulence}
We begin with a single-phase turbulence simulation to show that the forcing scheme is able to maintain a statistically stationary turbulent flow (simulation ``SP'' in table \ref{tab:TurbEmPhiVary}) and to compare it with results available in literature. A domain of $256^3$ lattice nodes representing a length $(2\pi)^3$ is initialized with a uniform initial density of $\rho^\alpha = 4.0$ [lu]. The relaxation time of the two fluids is set to $\tau = 0.5141$ which gives a viscosity of $\nu = 0.0047$ [lu] (\cite{perlekar2012droplet} use a similar value with $\tau=0.515$), which is a low enough viscosity to sustain turbulence while still being numerically stable. The forcing is concentrated around $k_f = 2\kmin$ and is distributed in the range of $k = \kmin$ to $8\kmin$. Further, $A=0.0005$, which generates a turbulent flow with a Taylor microscale of $\lambda = 13$ [lu], $Re_\lambda = 95$, $\tau_k = 97$ [lu], $\eta=0.7$ [lu] ($\kmax\eta = 2.2$) and $\ang{\epsilon}\approx 5\times10^{-7}$ [lu], which are calculated \textit{a posteriori}. The simulation is performed for $10^5$ $\Delta t$, which corresponds to $1000 \tau_k$. 

\begin{figure}
  \centerline{\includegraphics[width=0.75\linewidth]{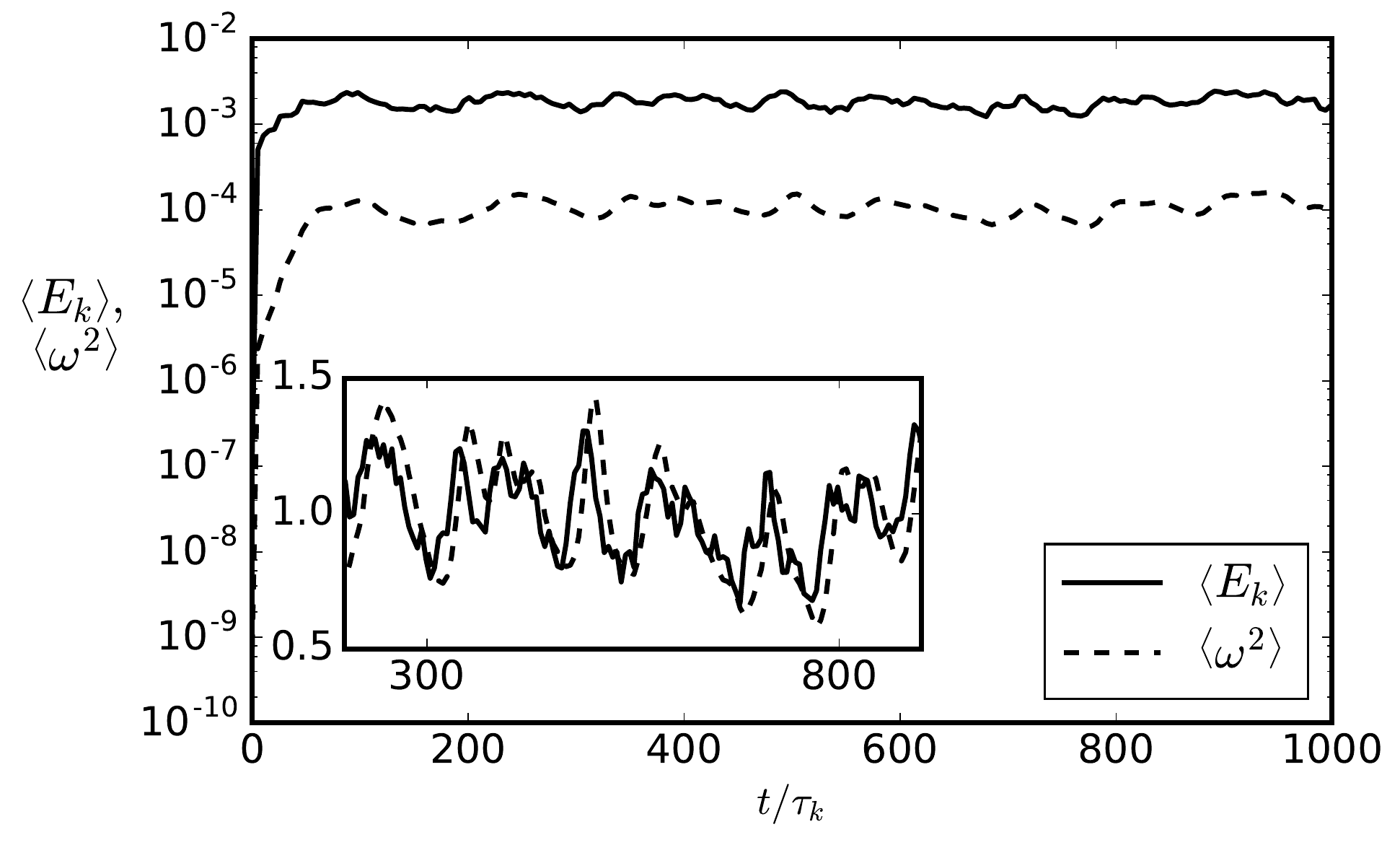}} 
  \caption{Evolution of average kinetic energy $\ang{E_k}$ and enstrophy $\ens$ in the single-phase turbulence simulation with $Re_\lambda = 95$. Both $\ang{E_k}$ and $\ens$ reach steady state confirming the balance between the energy dissipation and power input. In the inset, both profiles have been normalized by their time averaged value over the latter $3/4$th of the simulation duration.}
\label{fig:turbSinglePhaseKEevol}
\end{figure}

Figure \ref{fig:turbSinglePhaseKEevol} shows the evolution of $\ang{E_k}$ and $\ang{\omega^2}$ which attain their steady state values around $100\tau_k$ and continue to oscillate around this value. The crests and troughs of the $\ang{E_k}$ evolution show up in the $\ang{\omega^2}$ evolution with a slight delay (as seen in the inset of figure \ref{fig:turbSinglePhaseKEevol}, where the quantities have been normalized with their time averaged values over the latter $3/4$th of the simulation duration). This has been observed before, and ascribed to the energy cascading mechanism \citep{pearson2004delayed,biferale2011lattice} while \cite{tsinober2009informal} acknowledges this feature without invoking a cascade.

\begin{figure}
  \centerline{\includegraphics[width=\linewidth]{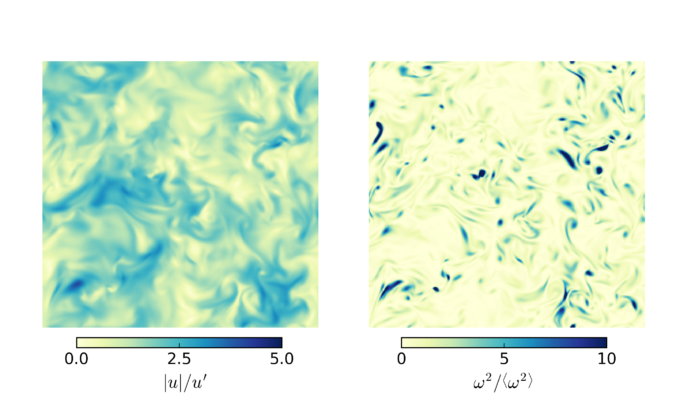}} 
  \caption{Cross-sections (at $z=N_x/2$) show snapshots of the velocity magnitude $|u|$ (left) and enstrophy $\omega^2$ (right), indicative of the rate of energy dissipation at time $t=500\tau_k$. Features typical of turbulent flow can be seen, where the velocity field shows features across several length scales while enstrophy remains localized in small scale structures.}
\label{fig:turbSinglePhaseFields}
\end{figure}

Figure \ref{fig:turbSinglePhaseFields} shows typical velocity and enstrophy field snapshots from a planar cross-section in the center of the domain at $500\tau_k$. The velocity field shows motions across various scales, while the enstrophy field (which is the square of the vorticity and relates directly to the rate of energy dissipation) shows typical small scale localized structures. Also note that $\omega^2$ assumes values as much as $10$ times the average $\ens$ (while at higher $Re_\lambda$, more extreme values are found), showing that intermittency is well reproduced in the simulations. This patchy structure of enstrophy (and hence dissipation) is an important factor to consider in simulations of turbulent dispersions, as the local rate of energy dissipation sets the local maximum stable droplet diameter.

\begin{figure}
  \centerline{\includegraphics[width=0.75\linewidth]{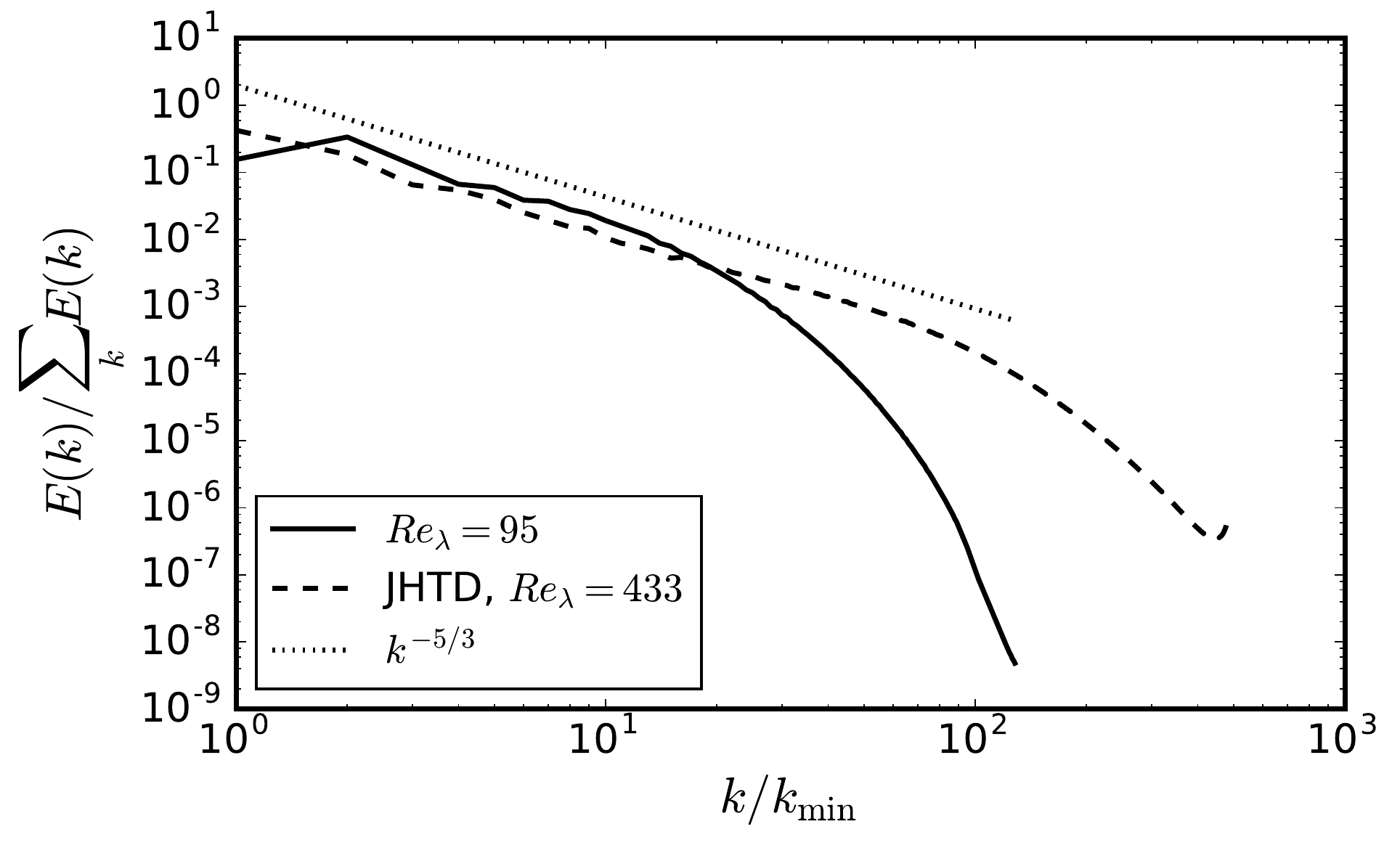}} 
  \caption{Kinetic energy spectrum for the single-phase simulation shown together with a sample spectrum from the Johns Hopkins Turbulence Database (JHTD, with $Re_\lambda = 433$). The chosen normalization is only to compare the shape of the two spectra along with a $k^{-5/3}$ inertial range scaling. The spectrum is further averaged over 20 realizations separated by $50\tau_k$.}
\label{fig:turbSinglePhaseSpectra}
\end{figure}

The kinetic energy spectrum is shown in figure \ref{fig:turbSinglePhaseSpectra}, along with a benchmark spectrum from the Johns Hopkins Turbulence Database \citep{li2008public} for a homogeneous isotropic turbulence simulation with $Re_\lambda = 433$ (on a grid of $1024^3$, generated with a spectral solver). The energy $E(k)$ has been normalized by the total energy $\sum_k E(k)$, and the wavenumber is normalized to show multiples of $\kmin$, which is done to compare the two spectra. A well developed inertial range is seen to exist, following the $k^{-5/3}$ spectral slope, which falls off around $k=30\kmin$ in our simulation. Lastly, in this simulation $u^\prime = 0.034$ [lu], and since the speed of sound is $c_s = 1/\sqrt{3}$ [lu], the flow Mach number is $M\!a = 0.06$ which is well within the incompressibility limit.

\section{Turbulent emulsions}\label{sec:turbulentEmulsions}
\subsection{Simulation setup}
The turbulent emulsion simulations are initialized with two fluids, which we denote by $\alpha$ (the carrier fluid) and $\beta$ (the droplet fluid). For a chosen volume fraction $\phi$ of fluid $\beta$, a single spherical droplet (a $\beta$-rich region) is initialized in the center of the domain which is otherwise $\alpha$-rich. The droplet density is denoted by $\rho_\beta^{\mathrm{in}}$, i.e the density of $\beta$ in the $\beta$-rich region, while $\rho_\beta^{\mathrm{out}}$ denotes the dissolved amount of component $\beta$ in the $\alpha$-rich region (i.e. the continuous phase), and likewise for component $\alpha$. Further, $\rho_\beta^{\mathrm{avg}}$ is used to refer to the average density of component $\beta$ in the entire domain. During the simulation, these density values can change to some extent depending on the $G_{\alpha\beta}$ parameter, though due to the symmetry of the model we have $\rho_\beta^{\mathrm{in}}/\rho_\alpha^{\mathrm{in}} \approx 1$ and $\rho_\beta^{\mathrm{out}}/\rho_\alpha^{\mathrm{out}} \approx 1$, which well represents many oil in water emulsions. We also keep $\nu_\beta/\nu_\alpha = 1$ (with $\nu = 0.0047$ [lu]). Spurious velocities ($u^{\mathrm{sp}}$) in these simulations have been limited to values sufficiently smaller than the physical velocity, so that their influence on the results is negligible. Typically, the large scale velocity $\mathcal{U}\sim O(10^{-2})$ while $u^{\mathrm{sp}}_{\mathrm{mean}}\sim O(10^{-4})$ and $u^{\mathrm{sp}}_{\mathrm{max}}\sim O(10^{-3})$ (for the range of $G_{\alpha\beta}$ values used in this study). Given that the speed of sound in these simulations $c_s = 1/\sqrt{3}$, we maintain that $ u^{\mathrm{sp}} \ll \mathcal{U} \ll c_s $, which is in line with our recent findings for emulsion droplets simulated with PP-LB \citep{mukherjee2018simulating,berghout2019simulating}.

We carried out three sets of simulations, the details of which are mentioned in table~\ref{tab:TurbEmPhiVary}. In all these simulations, the turbulence force is applied starting at $t=0$. The turbulence energy density $\ang{E_k}$ in an emulsion, for the same forcing amplitude, can be an order of magnitude lower than in single-phase turbulence. The Kolmogorov scale values have been calculated using the scaling $\eta \approx (\nu^3/\overline{\ang{\epsilon}})^{1/4}$ where $\overline{\ang{\epsilon}}$ is the spatio-temporally averaged dissipation rate (with $\overline{\ang{.}}$ denoting time averaging after the first quarter of the simulation time, during which the flow is well developed). We report $\eta$ upto two decimal places that follow from this scaling. The three sets are divided as follows

\begin{itemize}
\item Set 1 (P1-P5): In these simulations, only the dispersed phase volume fraction has been changed (from $\phi = 0.01$ to $\phi= 0.45$). Here $\eta$ is found to increase in simulations P1-P5, which is because the turbulence forcing scale $\mathcal{L}$ remains the same while $Re_\lambda$ decreases, hence reducing the separation between the largest and smallest scales.
\item Set 2 (T1-T5): In these simulations, the turbulence force amplitude is varied to change $Re_\lambda$ (at a fixed volume fraction $\phi=0.10$). For case T5, the interfacial tension has also been increased. Due to increasing $Re_\lambda$ in these simulations, since $\mathcal{L}$ is kept constant, $\eta$ is found (as expected) to decrease.
\item Set 3 (D1-D5): In these simulations, the domain size is increased while keeping the forcing lengthscale $\mathcal{L}$, amplitude and volume fraction ($\phi=0.15$) fixed, which keeps the turbulence energy density (or $Re_\lambda$) fixed. An additional simulation, D5, has been performed where the turbulence intensity and volume fraction have been increased for comparison with case D4. For all cases, $\eta$ remains almost constant as $Re_\lambda$ is kept constant by varying $\mathcal{L}$ (so that the ratio $\mathcal{L}/\eta$ is constant). In simulation D5, $Re_\lambda$ is increased fourfold in comparison to D1-D4, yet $\eta$ is the same as the increase in $Re_\lambda$ is achieved by the added scale separation due to a fourfold decrease in the forcing wavenumber in D5 ($k_f=1.5$) as opposed to D4 ($k_f=6.0$).
\end{itemize}

\begin{table}
  \begin{center}
\def~{\hphantom{0}}
  \begin{tabular}{lccccccccccccc}
  	 Sim & $N$ & $\nu$ & $G_{\alpha\beta}$ & $\phi$ & $A$ & $k_a,k_f,k_b$ & $\gamma$ & $\ang{E_k}$ & $\ang{\epsilon}$ & $\tau_k$ & $Re_\lambda$ & $\eta$ \\ [3pt]
  	 SP & $256^3$ & 0.0047 & - & - &0.0005 & 2.0 & - & $1.8\times 10^{-3}$ & $5.0\times 10^{-7}$ & 97 & 95 & 0.7\\
  	 \\
  	 P1 & $256^3$ & 0.0047 & 0.015 & 0.01 & 0.0005 & $1,2,8$ & 0.017 & $2.0\times 10^{-4}$ & $2.21\times 10^{-8}$ & $461$ & $51$ & $1.47$ \\
  	 P2 & $256^3$ & 0.0047 & 0.015 & 0.06 & 0.0005 & $1,2,8$ & 0.017 & $2.0\times 10^{-4}$ & $2.10\times 10^{-8}$ & $474$ & $53$ & $1.49$\\
  	 P3 & $256^3$ & 0.0047 & 0.015 & 0.15 & 0.0005 & $1,2,8$ & 0.017 & $1.7\times 10^{-4}$ & $1.93\times 10^{-8}$ & $493$ & $47$ & $1.52$ \\
  	 P4 & $256^3$ & 0.0047 & 0.015 & 0.2 & 0.0005 & $1,2,8$ & 0.017 & $1.5\times 10^{-4}$ & $1.75\times 10^{-8}$ & $518$ & $45$ & $1.56$\\
  	 P5 & $256^3$ & 0.0047 & 0.015 & 0.45 & 0.0005 & $1,2,8$& 0.017 & $1.3\times 10^{-4}$ & $1.65\times 10^{-8}$ & $534$ & $39$ & $1.58$ \\
  	 \\
  	 T1 & $256^3$ & 0.0047 & 0.015 & 0.10 & 0.00025 & $1,1.5,8$ & 0.017 & $8.4\times 10^{-5}$ & $4.90\times 10^{-9}$ & $980$ & $44$ & $2.14$\\
  	 T2 & $256^3$ & 0.0047 & 0.015 & 0.10 & 0.0005 & $1,1.5,8$ & 0.017 & $2.4\times 10^{-4}$ & $1.87\times 10^{-8}$ & $502$ & $64$ & $1.54$  \\
  	 T3 & $256^3$ & 0.0047 & 0.015 & 0.10 & 0.00075 & $1,1.5,8$ & 0.017 & $4.6\times 10^{-4}$ & $4.78\times 10^{-8}$ & $313$ & $78$ & $1.22$ \\
  	 T4 & $256^3$ & 0.0047 & 0.015 & 0.10 & 0.001 & $1,1.5,8$ & 0.017 & $6.5\times 10^{-4}$ & $8.36\times 10^{-8}$ & $237$ & $84$ & $1.06$ \\
  	 T5 & $256^3$ & 0.0047 & 0.016 & 0.10 & 0.001 & $1,1.5,8$ & 0.04 & $6.6\times 10^{-4}$ & $7.70\times 10^{-8}$ & $247$ & $90$ & $1.08$\\
  	 \\
  	 D1 & $128^3$ & 0.0047 & 0.015 & 0.15 & 0.0005 & $1,1.5,6$ & 0.017 & $1.2\times 10^{-4}$ & $1.80\times 10^{-8}$ & $511$ & $34$ & $1.55$\\
  	 D2 & $256^3$ & 0.0047 & 0.015 & 0.15 & 0.0005 & $1,3,6$ & 0.017 & $1.1\times 10^{-4}$ & $1.77\times 10^{-8}$ & $514$ & $30$ & $1.56$\\
  	 D3 & $384^3$ & 0.0047 & 0.015 & 0.15 & 0.0005 & $2,4.5,8$ & 0.017 & $1.1\times 10^{-4}$ & $1.81\times 10^{-8}$ & $509$ & $30$ & $1.55$\\
  	 D4 & $512^3$ & 0.0047 & 0.015 & 0.15 & 0.0005 & $3,6,9$ & 0.017 & $1.1\times 10^{-4}$ & $1.87\times 10^{-8}$ & $501$ & $30$ & $1.54$\\
  	 D5 & $512^3$ & 0.0047 & 0.015 & 0.2 & 0.0005 & $1,1.5,6$ & 0.017 & $4.2\times 10^{-4}$ & $1.80\times 10^{-8}$ & $511$ & $118$ & $1.55$\\
  	 
  \end{tabular}
  \caption{Simulations parameters for all cases. Here viscosity $\nu$ and interfacial tension $\gamma$ are in lattice units [lu], along with length and time measured as multiples of $\Delta x$ and $\Delta t$. The density and viscosity ratio between the components is kept at unity. The turbulence forcing is distributed over the range of wavenumbers from $k_a$ to $k_b$ centered at $k_f$. The fluid densities are initialized to $\rho_{\alpha,\beta}^\mathrm{in} = 4.0$  $\rho_{\alpha,\beta}^\mathrm{out} = 0.77$. The average kinetic energy $\ang{E_k} = (\sum_k E(k))/N$, and the average rate of energy dissipation $\ang{\epsilon}=(\sum_k 2\nu k^2E(k))/N$.}
  \label{tab:TurbEmPhiVary}
  \end{center}
\end{table}

To study the droplet characteristics in these simulations, we segment the droplets in space (also known as clustering) by thresholding the droplet density field at a cutoff value $\rho^c/\rho_\beta^{\mathrm{in}} = 0.57$ (which is effectively the density along the interface where $\rho^c \approx \rho_\alpha = \rho_\beta$) based on the algorithm used in \cite{siebesma2000anomalous}. This allows us to identify and mark all lattice points within individual droplets, which gives the droplet volume $V$, which in turn is used to calculate an effective diameter $d = (6V/\pi)^{1/3}$. Estimating the surface area of these droplets, which are in voxel form, requires more care. Often, the `GNU triangulation surface' (GTS) library \citep{popinet2004gts} is used in studies due to its efficient surface splitting operations (without the need for volumetric droplet segmentation). However, it was not used in this study as it did not provide a straightforward way of identifying droplets cut-off at domain edges due to periodicity (an issue implicitly resolved by our segmentation algorithm). Also, the GTS library was found to underpredict the surface area of a sphere by around $10\%$. Instead, we use the method proposed by \cite{windreich2003voxel} (originally developed for medical MRI data) to calculate surface area directly from voxels using a look-up table which divides surface voxels into 9 classes, and each class has a weighted contribution to the surface area. Using only the first 4 of these 9 classes, the area estimation error for a sphere was found to decrease to $1\%$, which was sufficiently accurate for our study.

\subsection{Effect of volume fraction}\label{sec:volFracVary}
\begin{figure}
  \centerline{\includegraphics[width=\linewidth]{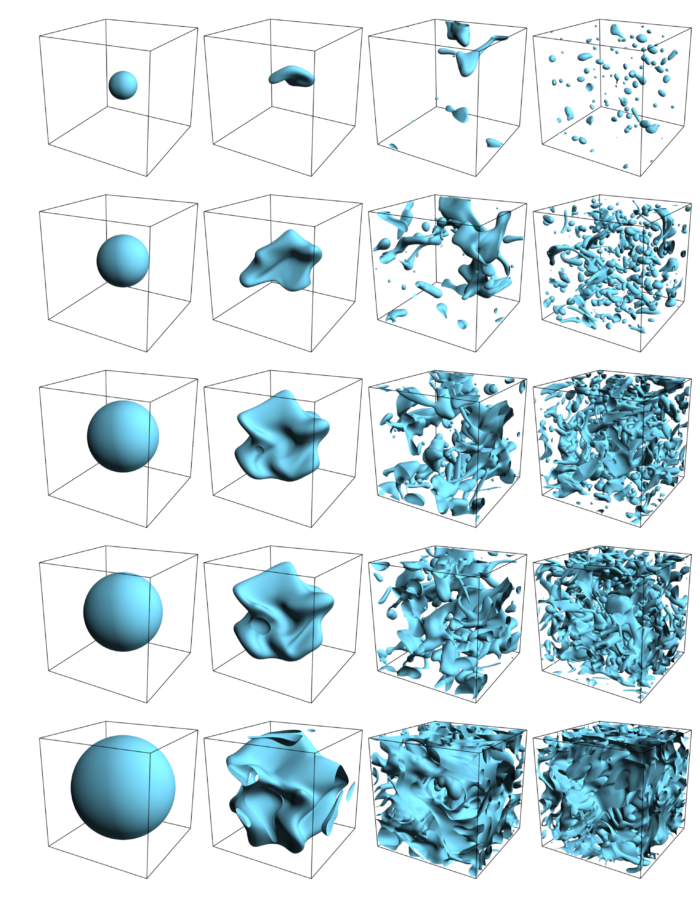}} 
  \caption{Dispersion formation under turbulence, for increasing volume fractions $\phi \in \left\lbrace 0.01,0.06,0.15,0.2,0.45\right\rbrace$ corresponding to simulation P1-P5 in table \ref{tab:TurbEmPhiVary} (top to bottom). The time instances are $t/\tau_k \approx 0, 10, 40, 100$ (left to right), and the dispersions are subjected to identical turbulence forcing.}
\label{fig:renderedPhiVaryTimeEvol}
\end{figure}

We now show results from simulations with varying dispersed phase volume fractions $\phi \in \left\lbrace 0.01,0.06,0.15,0.2,0.45\right\rbrace$ under identical turbulence forcing conditions (corresponding to P1-P5 in table \ref{tab:TurbEmPhiVary}). These simulations are performed for $10^5$ time steps. Figure \ref{fig:renderedPhiVaryTimeEvol} shows the dispersion formation process at various time instances starting from the initial spherical droplet of component $\beta$ shown as the iso-surfaces representing $\rho_\beta = \rho_\alpha$. The droplet begins to deform under the turbulent stresses, eventually breaking up to form a dispersion with a characteristic distribution.

Of the various volume fractions considered, $\phi=0.06$ and $0.15$ are most emulsion-like, i.e. they have a profusion of small droplets with a few large connected filaments. At $\phi=0.01$, the dispersed phase is too dilute to be considered an emulsion, although the droplet dynamics is interesting as the number of droplets $N_d$ and their characteristic diameters $d$ is small, and hence most of the droplets remain dispersed with relatively few coalescence events, and when droplets do coalesce, they break up soon after. At $\phi\geq 0.2$, most of the fluid volume remains connected, which is aggravated by the enhanced coalescence inherent to diffuse interface methods \citep{komrakova2015numerical,roccon2017viscosity}. This in turn is due to insufficient resolution of the interface with respect to the droplet sizes \citep{shardt2013simulations}, an effect we discuss more in depth in section \ref{sec:domainSize}. At higher turbulence intensity, the large connected regions can be expected to break into smaller droplets, and any coalescence will generate droplets of sizes larger than the maximum stable diameter, which will again breakup. 

\subsubsection*{Phase fraction evolution}
\begin{figure}
  \centerline{\includegraphics[width=0.75\linewidth]{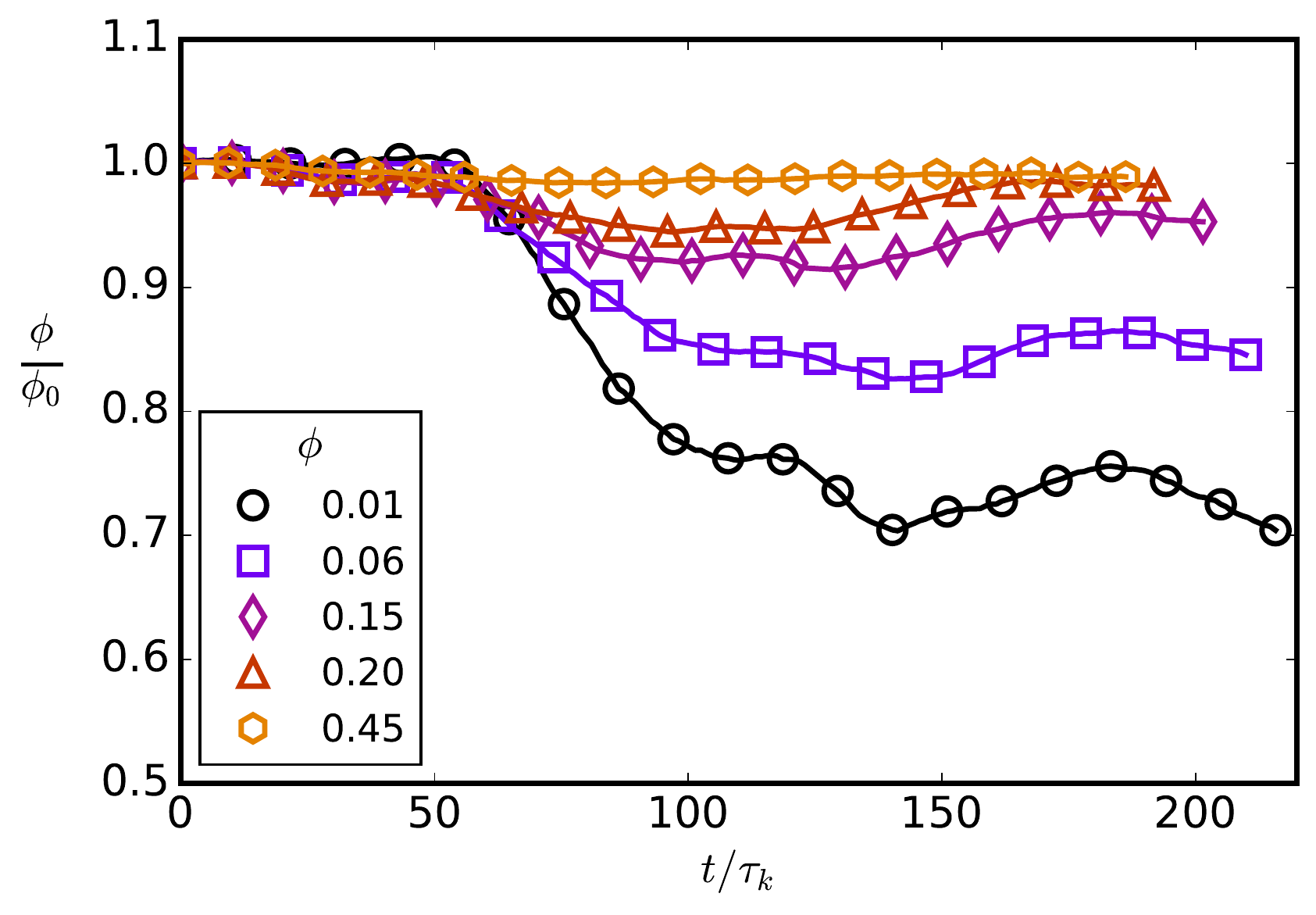}} 
  \caption{Evolution of volume fraction $\phi$ normalized by the initial volume fraction $\phi_0$, for the same turbulence intensity across simulations. There is more droplet dissolution for lower $\phi$ values, while the decrease is not monotonic as new smaller droplets can be formed as well.}
\label{fig:volFracVaryVolFracEvol}
\end{figure}

Figure \ref{fig:volFracVaryVolFracEvol} shows the evolution of the dispersed phase volume fraction $\phi$ normalized by the initial volume fraction $\phi_0$. There is a clear decrease over time (upto around $100\tau_k$) in the relative volume fraction, beyond which the value plateaus to a level around which it continues to oscillate (this will be confirmed subsequently from simulations T1-T5 in section \ref{sec:turbIntensity} which were performed for a five times longer duration). This relative reduction in $\phi$ is more pronounced at lower $\phi$ values (up to around $30\%$) than at higher $\phi$ (around $2-5\%$). Note that this is \textit{not} a mass conservation issue, as the total component mass is perfectly conserved in the system, and only the amount of component $\beta$ present as the dispersed phase reduces, which gets dissolved in the $\alpha$-rich (continuous phase) region. This is also why the relative decrease in $\phi$ is strongest for $\phi=0.01$, as the dissolution of $\beta$ into the continuous phase is provided by a very low number of droplets.

The reason for the reduction in $\phi$ is twofold. First is the dissolution of small droplets due to a finite interface width, which is an issue inherent to most diffuse interface methods. This can be characterized by the Cahn number $Ch$. Hence, if $Ch\sim O(1)$ (or greater), the droplet becomes unstable and is prone to dissolution. On the other hand, \cite{shardt2013simulations} showed for droplet collision in shear flow that coalescence is inhibited with decreasing $Ch$ number. In the limit of $Ch \to 0$, coalescence would cease to occur, while at larger $Ch$ numbers, coalescence is favoured. These considerations mandate having a finite $Ch$ number in the range $0< \ll Ch \ll 1$(for all droplet sizes in the system) for achieving steady state simulations while allowing for both coalescence and breakup. The second reason for the reduction in $\phi$ is its sensitivity to the segmentation threshold. In appendix \ref{app:Threshold} we demonstrate that \textit{only} this result, i.e. the evolution of the volume fraction, depends on the choice of the segmentation threshold. Part of the droplet phase fraction goes into constituting the increased interfacial region (i.e. roughly the total surface area of all droplets $S_A$ multiplied by the interface width $\dint$). Slightly varying the segmentation threshold to lower values (so that it is closer to $\rho_\beta^{\mathrm{out}}$), the volume fraction loss is reduced (which may indicate that $\rho^{\mathrm{c}} \neq (\rho^{\mathrm{out}} + \rho^{{in}})/2$), although the exact choice of $\rho^{\mathrm{c}}$ does not change our results. Further, the reduction in $\phi$ is also not monotonic, as mass of component $\beta$ dissolved in the $\alpha$-rich region can eventually accumulate inside other droplets. 

Droplet dissolution can be a debilitating numerical issue, where for instance \cite{biferale2011lattice,perlekar2012droplet} had to resort to artificially inflating droplets to maintain a constant phase fraction and \cite{komrakova2015numerical} reported that they could not attain steady state simulations with the free-energy LB method at low volume fractions as all droplets dissolved away into the continuous phase. In our PP-LB simulations, this issue is due to an interplay of three main factors - (i) the liquid-liquid repulsion $G_{\alpha\beta}$ which keeps the two components demixed, (ii) the turbulence intensity which breaks large droplets into smaller ones and (iii) the phase fraction which at low values makes $\rho_\beta^{\mathrm{out}} \approx \rho_\beta^{\mathrm{avg}}$ (i.e. at low $\phi$, phase segregation can become weaker). Despite being present, droplet dissolution is limited to a minor effect in our simulations. More precisely, the PP-LB method employed in this study can be used to reasonably simulate certain regions of the turbulent emulsions parameter space where droplet dissolution is not significant. Namely, for a given turbulence intensity ($Re_\lambda$), there will be a critical lower bound on the interfacial tension $\gamma_c$ such that droplets with $\gamma>\gamma_c$ can be simulated. For increasing $Re_\lambda$, $\gamma_c$ would increase as well, and its exact dependence on $Re_\lambda$ could be investigated by numerically mapping the phase space which is out of the scope of the current study. Similarly, there will be a lower bound on the value of $\phi$, below which all droplets will dissolve due to weak phase segregation when $\rho_\beta^{\mathrm{out}} \approx \rho_\beta^{\mathrm{avg}}$. Considering these related effects, we restrict ourselves to a parameter range where we can attain long, stable simulations to collect meaningful statistics pertaining to the droplet coalescence and breakup equilibrium. 

\begin{figure}
  \centerline{\includegraphics[width=\linewidth]{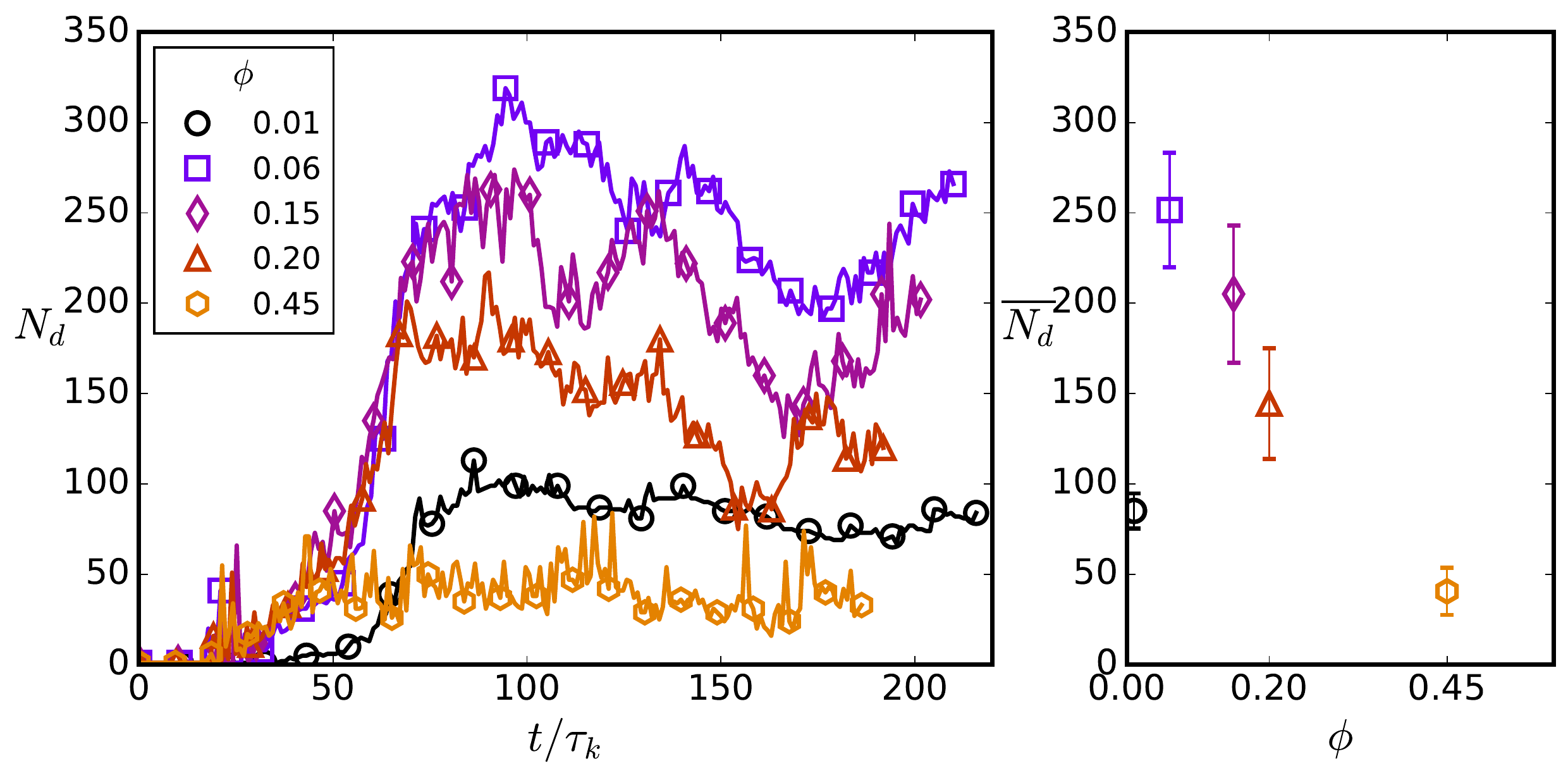}} 
  \caption{Evolution of the number of droplets ($N_d$) in the system, which attains its characteristic value around $75\tau_k$ and oscillates around a temporal mean. The $\phi=0.06$ case produces the highest number of droplets (around 250), which is seen on the right panel where $\overline{N_d}$ is $N_d$ averaged from $75\tau_k$ to $200\tau_k$, and the error bars show the standard deviation.}
\label{fig:volFracVaryNdEvol}
\end{figure}

\subsubsection*{Droplet number density evolution}
Figure \ref{fig:volFracVaryNdEvol} shows the evolution of the number of droplets ($N_d$) in the system for varying $\phi$. $N_d$ begins to increase following the first breakup events around $25\tau_k$ and rises steadily to its characteristic value around $75\tau_k$, around which it continues to oscillate. These oscillations in $N_d$ are indicative of competing coalescence and breakup dynamics. The falls in the $N_d$ evolution profiles are due to coalescence events, which generate droplets of large sizes that are unstable. These droplets then break up under turbulent stresses and $N_d$ increases again. Breakup is delayed for $\phi=0.01$ as compared to the other cases and $N_d$ only begins to increase around $50\tau_k$. This is because the size of initial droplet is much smaller ($\sim 64$ [lu]) than the forcing wavelength ($\sim 128$ [lu]), and the droplet starts to advect initially, as seen from figure \ref{fig:renderedPhiVaryTimeEvol}. When smaller scales are generated (around $50\tau_k$, as can be seen from the enstrophy evolution in figure \ref{fig:turbSinglePhaseKEevol}), the droplet begins to shear and break. The evolution of $N_d$ does not show large fluctuations for $\phi=0.01$ due to relatively fewer coalescence and breakup events in this case, which is because the droplets are smaller and more distant from each other than in higher $\phi$ cases.

Although $\phi = 0.15$ and $0.2$ simulations have a larger volume of fluid $\beta$, the number of droplets generated is lower than $\phi=0.06$. This is because of a higher propensity for coalescence in these systems which generates large connected regions and smaller satellite droplets. This is most prominently seen for $\phi=0.45$, where $N_d$ is even lower than $\phi=0.01$, as most of the fluid forms extended filaments that remain connected across the periodic boundaries. Increasing the turbulence intensity can be expected to generate more droplets at higher $\phi$, and hence for a given $Re_\lambda$, there will be a specific $\phi$ that maximizes the number of droplets formed and hence produce a more emulsion-like droplet size distribution.

\subsubsection*{Droplet size distribution}
\begin{figure}
  \centerline{\includegraphics[width=0.9\linewidth]{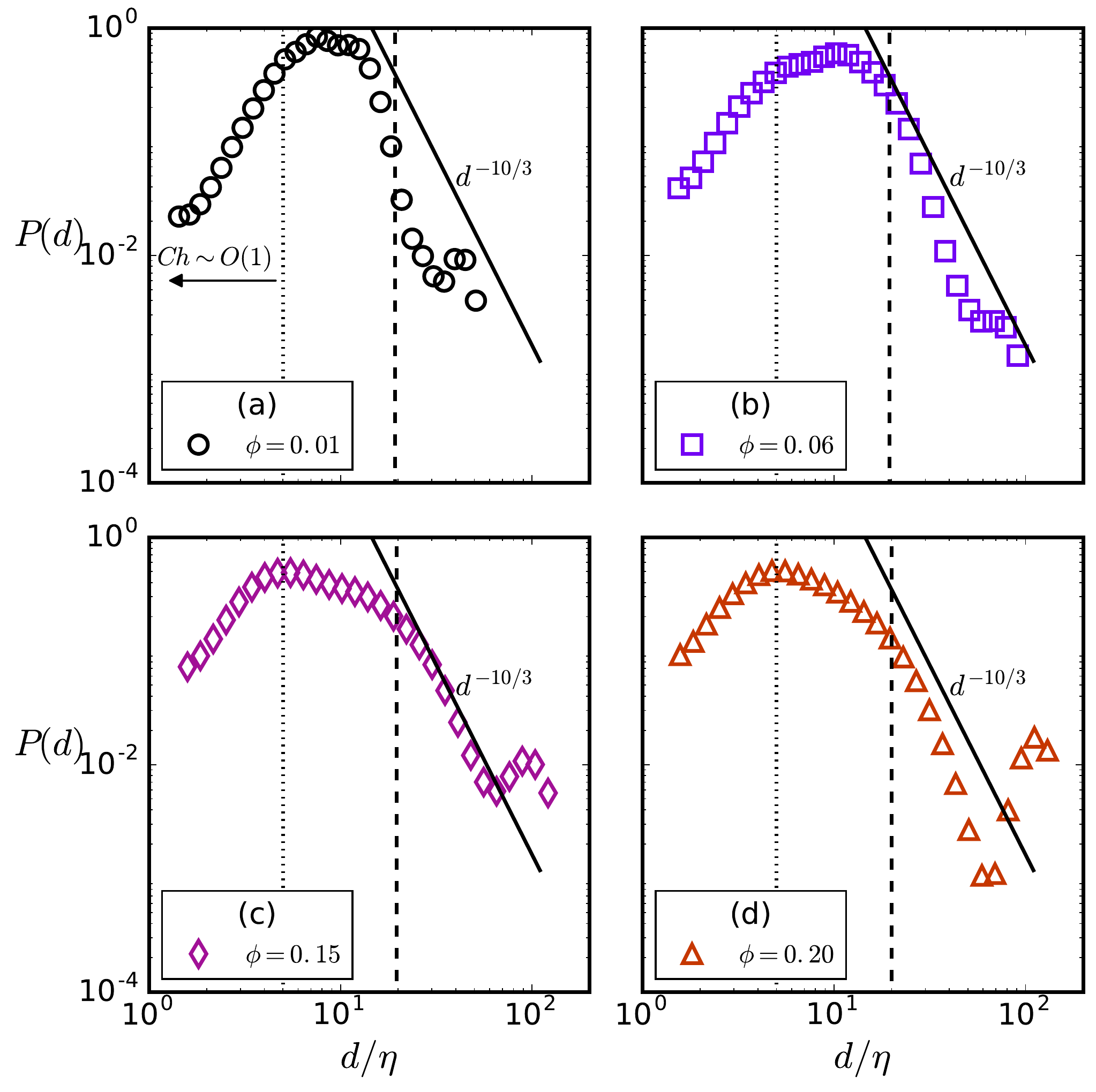}}
  \caption{Droplet size distributions for varying volume fractions (on a $256^3$ domain). Case (a) $\phi=0.01$ shows a peak around $d/\eta \approx 10$, which rapidly falls off at higher $d/\eta$. Cases (b) $\phi=0.06$ and (c) $\phi=0.15$ have a wider range of droplet sizes, and the distribution follows a $d^{-10/3}$ scaling in an intermediate $d/\eta$ range. For $\phi \geq 0.20$, a significant secondary peak at high $d/\eta$ indicates the few large connected regions that form in the periodic simulation domain, along with multiple smaller satellite droplets. The vertical dashed line shows the Hinze scale and the vertical dotted line marks the limit to the left of which the Cahn number $Ch\sim O(1)$ and droplets become unstable.}
\label{fig:volFracVaryDropDistrLog}
\end{figure}

Figure \ref{fig:volFracVaryDropDistrLog} shows the distribution of the equivalent droplet diameter $d = (6V/\pi)^{1/3}$ (where $V$ is the droplet volume) for varying $\phi$ (calculated with $25000-35000$ droplets identified between times $75\tau_k$ to $200\tau_k$, sampled at each $\tau_k$). Case (a) $\phi=0.01$  shows a peak around $d/\eta \approx 10$, beyond which the distribution rapidly falls off due to the dispersion being dilute (see 4th panel in the top row of figure \ref{fig:renderedPhiVaryTimeEvol}). Due to infrequent coalescence, large droplets are not formed very often. This was also reflected in the $N_d$ evolution (figure \ref{fig:volFracVaryNdEvol}) which does not fluctuate as much as higher $\phi$ simulations. Cases (b), (c) and (d) with $\phi \geq 0.06$ follow a $d^{-10/3}$ power law in an intermediate droplet size range $15<d/\eta<30$, showing the formation of larger droplets. This is in accordance to the prediction of \cite{garrett2000connection} and \cite{deane2002scale} for droplets in the inertial range of turbulence, which was also found by \cite{skartlien2013droplet} in their simulations.

Also, for $\phi\geq 0.15$, a secondary peak appears at high $d/\eta$, which is due to a few large connected regions forming due to coalescence, which remain connected despite occasional satellite droplets breaking off. Hence in these simulations, droplets in an intermediate range are less frequent, as upon formation they would soon coalesce with the larger connected region. This is first a consequence of having a high volume fraction at a lower turbulence intensity. At higher $Re_\lambda$, the large region would be unstable and hence break apart forming droplets with a range of diameters. Secondly, the formation of this larger connected region also depends on $Ch$. If a simulation is performed on a much larger domain for the same volume fraction $\phi=0.20$ and turbulence intensity $Re_\lambda=45$, due to an increased separation between $d$ and $\dint$ (lower $Ch$), coalescence would be inhibited. It should be noted that the uncertainty in determination of $d$ is around $10\%$, as shown in appendix \ref{app:Threshold}. 

Further, $\eta \approx 1.5$ [lu] here and given that the interface width $\dint \approx 5-6$ [lu], the $Ch$ for these droplets is approximately in the range $0.03 < Ch < 1.5$. Droplets towards the higher side where $Ch\sim O(1)$ will be unstable and prone to dissolution, which is reflected in the distributions falling off to the left of $d/\eta \approx 5$. In physical systems, small droplets are stable and can only be destroyed by coalescence. Resolving droplets in this range of diameters (where $d/\eta \sim O(1)$) will require over-resolving the Kolmogorov scale (to decrease the relative $Ch$), as was done by \cite{komrakova2015numerical}. Lastly, the length scales are ordered as $N_x > \mathcal{L} \gg d \gg \dint > \eta$ for cases P1-P3 while $N_x > \mathcal{L} > d \gg \dint > \eta$ for cases P4 and P5 (where due to higher $\phi$, the long droplet filaments can be of length $\sim \mathcal{L}$).

\begin{figure}
  \centerline{\includegraphics[width=\linewidth]{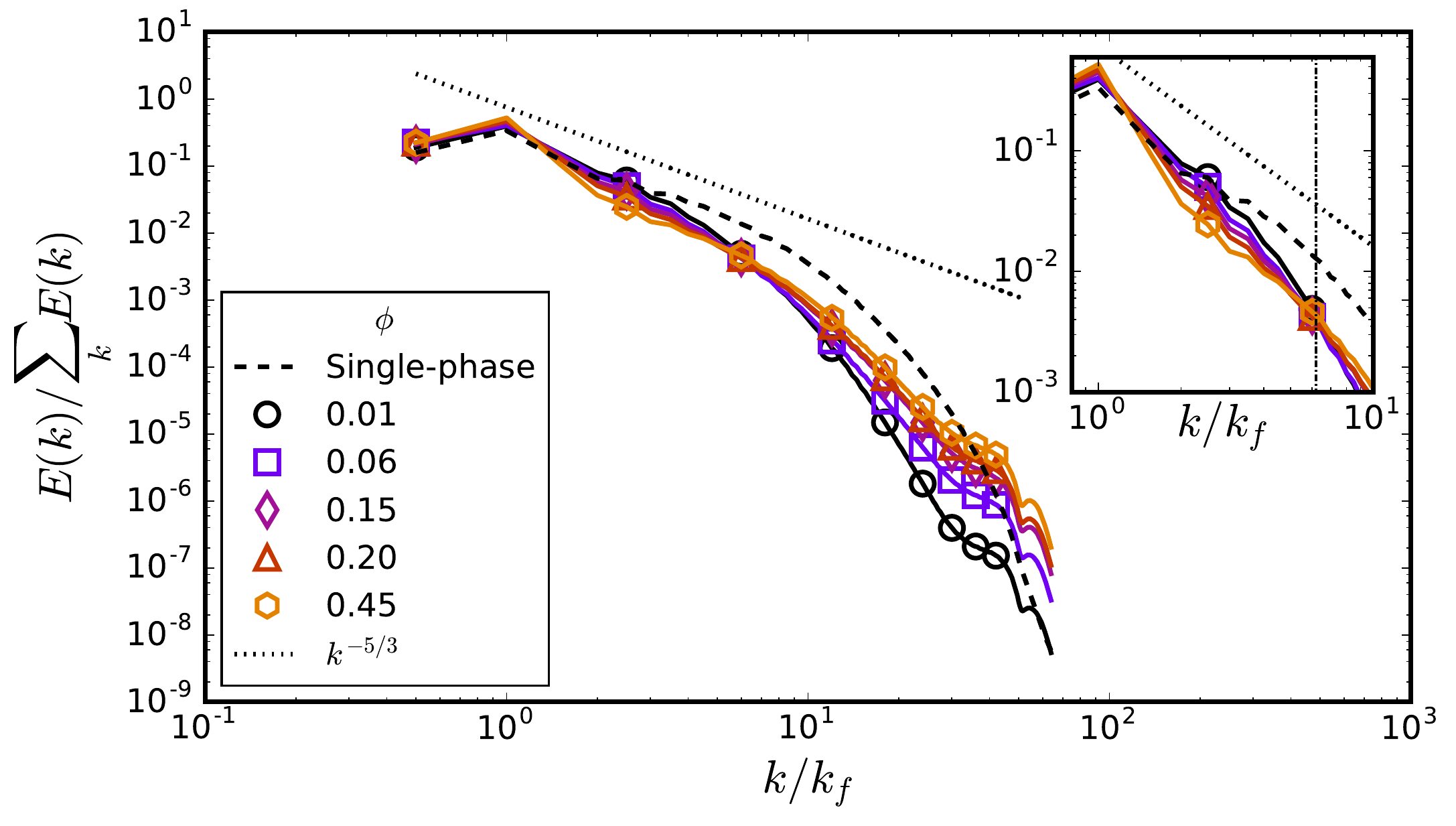}} 
  \caption{Kinetic energy spectra obtained from varying $\phi$ simulations (averaged between $75\tau_k$ and $100\tau_k$, sampled every $2\tau_k$). At higher $\phi$ values, the turbulence cascade is suppressed at intermediate wavenumbers (seen as deviations from Kolmogorov's $k^{-5/3}$ scaling). At higher wavenumbers, there is an inverse energy cascade due to droplet coalescence which adds kinetic energy to the smaller scales, which is stronger at higher $\phi$ values due a higher chance of coalescence in a dense dispersion. The vertical line in the inset figure corresponds approximately to the inverse of the Hinze scale.}
\label{fig:volFracVaryKESpectra}
\end{figure}

\subsubsection*{Multiphase kinetic energy spectra}
Figure \ref{fig:volFracVaryKESpectra} shows the kinetic energy spectra for the droplet laden simulations, in comparison to the single-phase turbulence simulation with identical forcing. The first effect to note is the suppression of the inertial range (i.e. deviation from the $k^{-5/3}$ law) which is seen more clearly in the inset figure, and has been found previously \citep{perlekar2014spinodal}. For increasing $\phi$, the spectra between $1<k/k_f<10$ shift away from the inertial range scaling and the single-phase spectrum, which shows that the cascading mechanism becomes weaker. Interestingly, the spectra pass through a single point, which is marked by the vertical line in the inset figure. This point is very close to the inverse of the Hinze length scale given by $d_{\mathrm{max}} = 0.725 (\rho/\gamma)^{-3/5}\epsilon^{-2/5}$. Beyond this point, the higher $\phi$ simulations contain higher energy at the smaller scales (large wavenumbers). This is due to coalescence which generates small scale eddies and is more frequent at higher $\phi$. Two or more droplets coalescing add kinetic energy to the flow by loss of surface energy due to a reduction in overall surface area. The crossover of the multiphase spectra (for $\phi\geq 0.15$ cases) with the single-phase spectrum shows that the dissipation range has higher energy in the presence of droplets, as was also reported by \cite{perlekar2014spinodal}. Interestingly, \cite{ten2004fully} also found such a spectral crossover at increasing volume fractions for solid spherical particles in turbulence. 

The $\phi=0.01$ simulation has the lowest energy at high wavenumbers, as coalescence events are rare, and the droplet sizes are smaller which derive energy from eddies corresponding to higher wavenumbers. Lastly, a small jump in the spectra at $k/k_f \approx 50$ is consistently seen for all cases, which corresponds precisely with the interface width in our simulations (i.e $5-6$ [lu]). The extra energy there is due to the spurious currents present in the system, which are found to be much weaker that the physical velocity scales. \cite{komrakova2015numerical} reported that spurious currents completely dominated the higher wavenumbers of the kinetic energy spectra in their turbulent dispersion simulations, due to which the spectra could not be well represented. Our work does not suffer from this problem, and although spurious currents are present, they do not adversely influence our results.

\subsection{Effect of turbulence intensity}\label{sec:turbIntensity}
\begin{figure}
  \centerline{\includegraphics[width=0.9\linewidth]{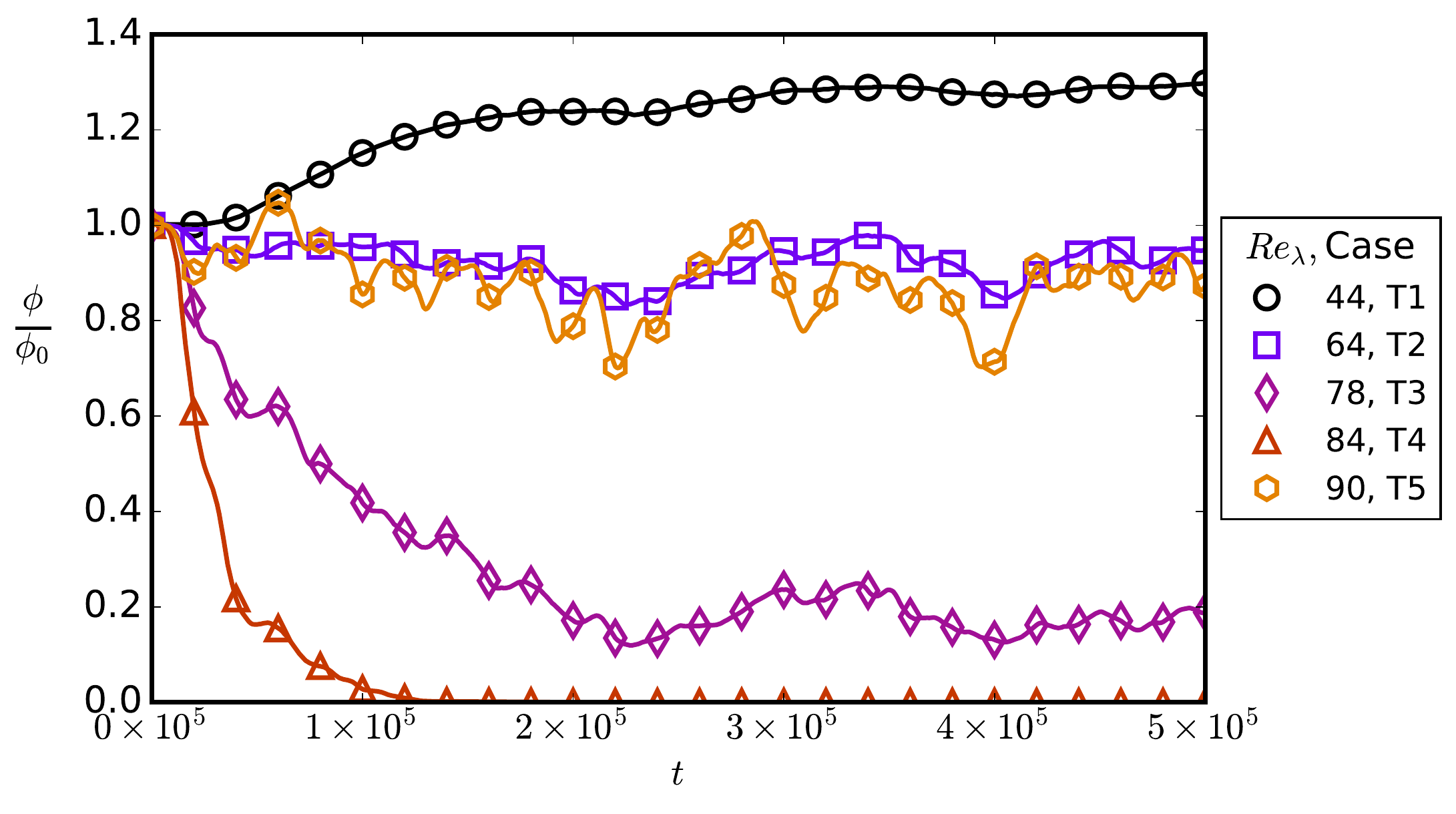}} 
  \caption{Evolution of the relative volume fraction $\phi/\phi_0$ for varying turbulence intensity simulations (cases T1-T5 in table \ref{tab:TurbEmPhiVary}). Increasing $Re_\lambda$ causes greater droplet dissolution leading to a lower settling value of $\phi/\phi_0$. This effect limits the parameter space that can be simulated with the original PP-LB method, as shown by cases T4 and T5.}
\label{fig:TurbVaryVolFracEvol}
\end{figure}

As mentioned earlier, the idea behind applying turbulence is to cause fragmentation of the dispersed phase, and the number of droplets thus formed depends upon the intensity of turbulence. We now keep the volume fraction fixed at $\phi=0.1$ and increase the turbulence intensity by increasing the forcing amplitude. These are simulations T1-T5 in table \ref{tab:TurbEmPhiVary}, and are run for $t=0.5$ million time steps each, though the simulations will have different $\tau_k$. Figure \ref{fig:TurbVaryVolFracEvol} shows the evolution of the normalized phase fraction over time, and as expected, at higher turbulence intensities (which leads to a higher $Re_\lambda$), $\phi/\phi_0$ reduces over time to an individual stable value. For the case T4, all the droplets dissolve within $10^5\ \Delta t$ , which shows that for this combination of parameters (refer to table \ref{tab:TurbEmPhiVary}), turbulence forcing undesirably outclasses the PP-LB phase segregation. The small droplets formed in this system are subsequently unstable (due to $Ch\sim O(1)$), which causes complete dissolution of the dispersed phase. Upon increasing the liquid-liquid repulsion parameter $G_{\alpha\beta}$ (hence also changing the fluid composition and dimensionless numbers that include interfacial tension, like the Weber or Ohnesorge number) in case T5, we see that for the same turbulence intensity as case T4, $\phi/\phi_0$ remains stable. This reaffirms that, with the original PP-LB method, certain regions of the turbulent emulsions parameter space can be simulated properly, while in other cases (case T4 and to some degree also case T3) simulations may require additional numerical remedies like the mass correction scheme of \cite{biferale2011lattice,perlekar2012droplet} or an enhanced Kolmogorov scale resolution (to achieve higher Cahn numbers) as done by \cite{komrakova2015numerical}.

\subsubsection*{Droplet number density evolution}
\begin{figure}
  \centerline{\includegraphics[width=0.9\linewidth]{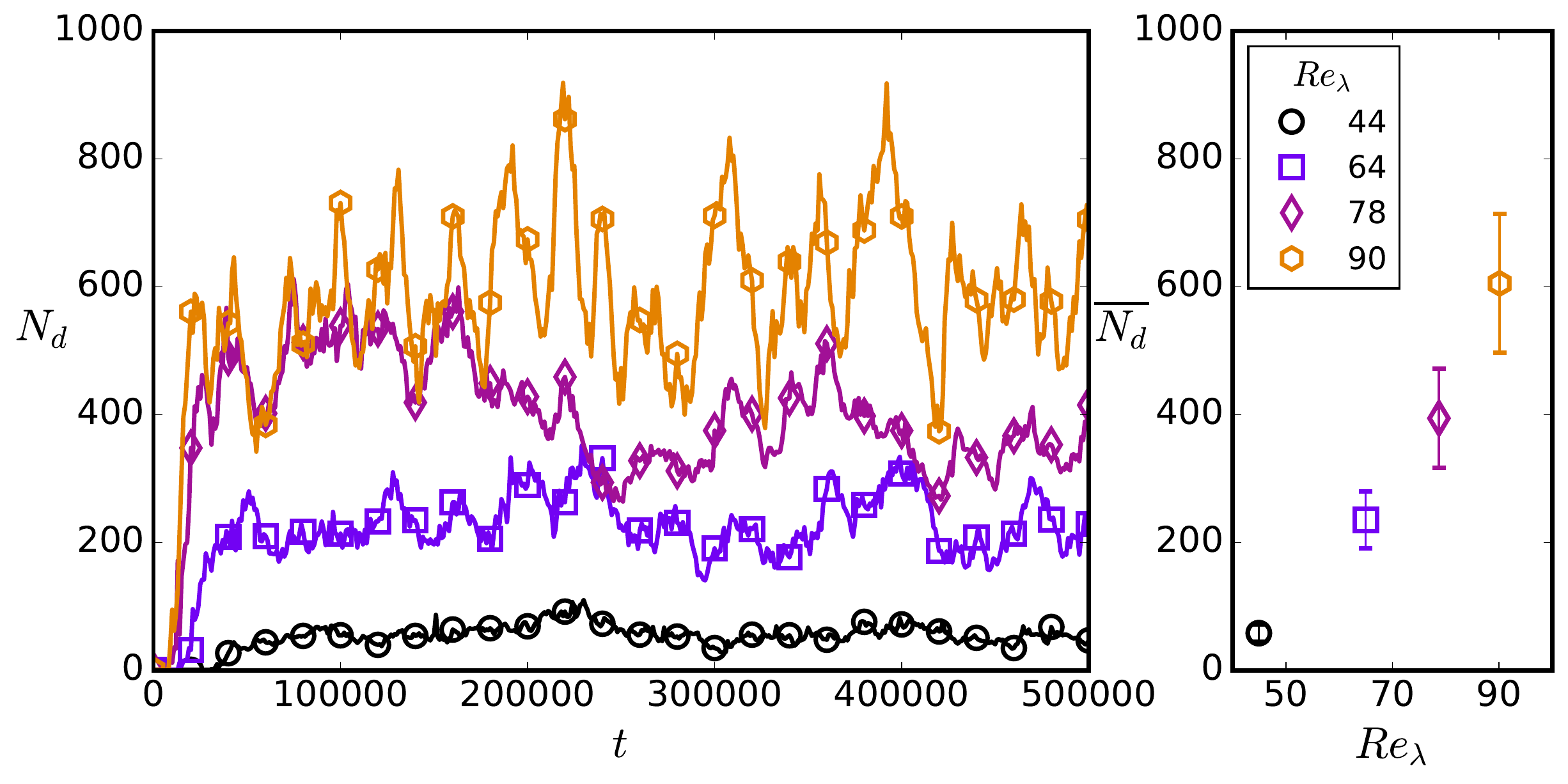}} 
  \caption{Evolution of the number of droplets ($N_d$) for increasing turbulence intensity, indicated by $Re_\lambda$. Increasing $Re_\lambda$ leads to a larger number of droplets in the system, also widening the droplet distributions as seen from the fluctuations in the $N_d$ evolution. Here $N_d$ is averaged between times $t=1\times 10^5$ and $t=5\times 10^5$.}
\label{fig:TurbVaryNumDropEvol}
\end{figure}

Figure \ref{fig:TurbVaryNumDropEvol} shows the evolution of the number of droplets for cases T1-T5 for varying turbulence forcing amplitudes (excluding case T4 where all the droplets eventually dissolve due to a relatively weaker inter-component repulsion). Increasing $Re_\lambda$ increases the average number of droplets in the system (obtained by averaging $N_d$ between times $t=1\times 10^5$ and $t=5\times 10^5$) from around $\overline{N_d}=50$ for $Re_\lambda=44$ to $\overline{N_d} = 600$ for $Re_\lambda=90$. Further, two interesting features in the evolution of $N_d$ can be noted. First is that the variation in $N_d$ increases with $Re_\lambda$, which results in a larger standard deviation of $\overline{N_d}$. This also makes it possible to generate a wider distribution of droplet diameters in the system. The second striking feature is the quasi-periodic rise and fall in the droplet number concentration (with a period of around $8-10\tstar$), most distinctly seen for the $Re_\lambda = 90$ simulation (case T5). There seems to be an upper limit to the number of droplets that can be formed, which apart from constraints of resolution and maximum sphere-packing of the domain while keeping the diffuse interfaces apart, indicates also at the underlying physical mechanisms. At its peak, $N_d \approx 900$ here, a state corresponding to most droplets being rather small that cannot undergo additional breakup as they would all be well below the Hinze scale. These droplets are advected around by the flow, and they begin to coalesce when they collide, causing $N_d$ to drop to its lower limit, where a significant number of droplets will again be larger than the Hinze scale, and they begin to break and this cycle continues. We shall revisit this feature in detail in section \ref{sec:limitCycle}.

\subsubsection*{Dispersion morphology}
The dispersion morphology can be quantified with the concentration spectrum $k^2 S(k,t)$, a quantity commonly used to describe coarsening dynamics (or spinodal decomposition) \citep{chen2000ternary,perlekar2014spinodal}. Here $S(k,t)$ is the shell-averaged structure factor which is obtained using the Fourier transform $\hat{\phi}_\mathbf{k}$ of the density-density correlation function $\phi - \overline{\phi}$, where $\phi = (\rho_\alpha - \rho_\beta)$ and $\overline{\phi}$ is the mean value of $\phi$. The quantity $\hat{\phi}_\mathbf{k}$ is shell-averaged in wavenumber space to obtain $S(k,t)$ as follows
\begin{equation}
S(k,t) = \frac{\sum_k |\hat{\phi}_\mathbf{k}|^2}{\sum_k 1}
\end{equation}
Here $\sum$ denotes summation over wavenumber shells $k\in [k-1/2, k+1/2]$ where $k = \sqrt{\mathbf{k} \cdot \mathbf{k}}$. Further, a characteristic length $L(t)$ can be calculated using the first moment of $S(k,t)$ as follows
\begin{equation}
L(t) = 2\pi\frac{\sum_k S(k,t)}{\sum_k kS(k,t)}
\end{equation}

\begin{figure}
  \centerline{\includegraphics[width=0.85\linewidth]{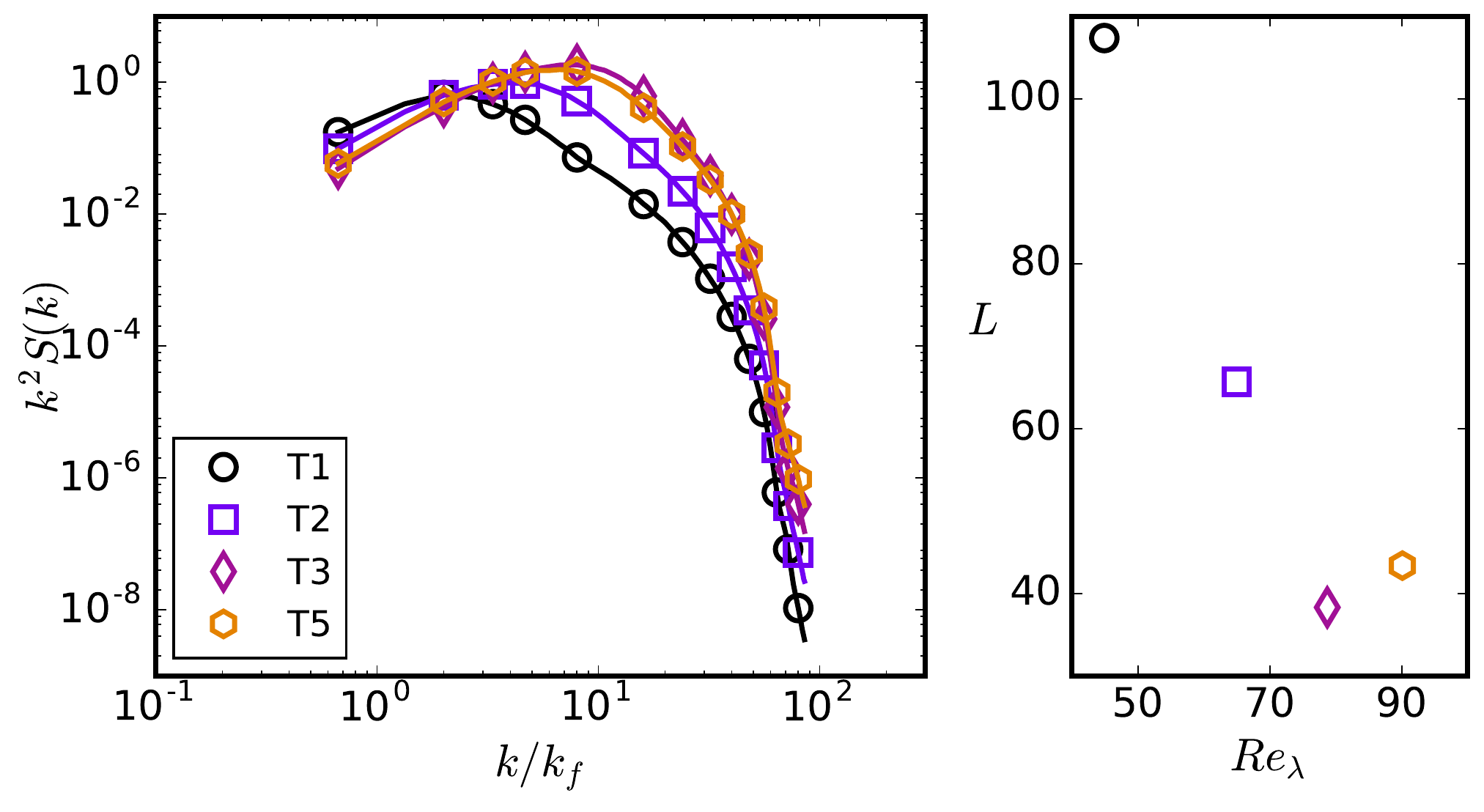}} 
  \caption{Concentration spectrum and characteristic length characterizing the dispersion morphology for increasing turbulence intensity simulations, corresponding to cases T1-T5 in table \ref{tab:TurbEmPhiVary}. The structure factor $S(k,t)$ was time averaged over 10 realizations separated by $\approx 50\tau_k$, and further normalized by $\sum_k S(k,t)$ to compare the relative difference in concentration at each wavelength. Increasing $Re_\lambda$ generates smaller droplets which is seen in the reduction of the characteristic length.}
\label{fig:TurbVaryStructureFactor}
\end{figure}

Figure \ref{fig:TurbVaryStructureFactor} shows the concentration spectrum for cases T1-T5, which reveals the influence of the turbulence intensity on the dispersion morphology. As $Re_\lambda$ is increased, smaller droplets begin to dominate the system which is seen from the shift towards higher wavenumbers in $k^2S(k,t)$. This is also reflected in the time averaged characteristic length $L$ which decreases from $100$ to around $40$ [lu]. Note that cases T3 and T5 almost overlap. This shows that the turbulence intensity and the repulsion parameter $G_{\alpha\beta}$ (or interfacial tension) compete to create a particular morphology, and a similar droplet distribution may be attained by varying the two factors in tandem.

\subsection{Effect of domain size}\label{sec:domainSize}
In simulations corresponding to D1-D4 in table \ref{tab:TurbEmPhiVary}, we successively increase the domain size $N_x$ while keeping the turbulent energy density the same. This essentially creates a separation between the domain size $N_x$ and the forcing scale $\mathcal{L}$, and allows for a better resolution of the largest droplet extension before breakup. So far, studies on turbulent dispersions have focused on maximizing the turbulence intensity which is reflected in the general proclivity for achieving higher $Re_\lambda$ in DNS simulations with Lagrangian objects like particles or droplets \citep{toschi2009lagrangian}. This finds implicit justification in that $Re_\lambda$ in real systems where droplets and turbulence interact is typically very high (for instance droplet-turbulence interaction in clouds occurs at $Re>10^6$ \citep{falkovich2002acceleration,shaw2003particle}, where $Re_\lambda = \sqrt{15 Re}$ for homogeneous, isotropic turbulence). In periodic domain DNSs, a high $Re_\lambda$ is achieved by minimally resolving the Kolmogorov scale (the $k_\mathrm{max}\eta > 1$ condition \citep{moin1998direct}), while forcing turbulence at the largest possible scales i.e. $\mathcal{L} \approx N_x$ or $k_f \approx 1-2 \kmin$. This wide separation of scales manifests a high $Re_\lambda$. There are a few connected issues regarding the relative resolution of the various length scales, which is the focus of this section.

The first issue, emphasized by \cite{komrakova2015numerical}, is the utility of over-resolving the Kolmogorov scale ($\eta \approx 10$ as opposed to $1$ [lu]), which helped remedy the rapid dissolution of droplets in their simulations. The increased resolution of $\eta$ and $d$ can also be seen as a reduction in the size of the interface $\dint$, i.e. an decrease in the Cahn number $Ch$, since the interface thickness (in terms of the \textit{number} of lattice spacings) remains constant while smaller droplets and turbulent length scales become better resolved (i.e. they become larger relative to $\dint$). Droplet dissolution also depends on the relative strengths of turbulence and phase segregation (effectively the interfacial tension), as was demonstrated in section \ref{sec:turbIntensity}. 

The other issue is that weak large scale forcing introduces a caveat that droplets tend to deform into long, slender filaments that stay connected across the periodic domain. The length scale of the largest droplet extension before breakup $d^{\mathrm{ext}}$ can become comparable to $N_x$, which means that breakup cannot be resolved. The dispersion then forms a complex tangled structure, which does not morphologically resemble an emulsion. This issue is aggravated by high volume fractions of the dispersed phase.

In simulations D1-D4, we increase the forcing wavenumber $k_f$ by the same factor as the domain size $N_x$ (while keeping the forcing amplitude $A$ the same). The upper and lower wavenumber bounds ($k_a, k_b$) are also suitably adjusted to distribute the forcing over a reasonable wavenumber range (and all integer values in the range $k\in \left[k_a,k_b \right]$ are considered). This ensures that the energy density remains the same in these simulations, while larger droplet deformations ($d^{\mathrm{ext}}$) can be resolved accurately. Successively increasing the domain size in this way allows separating $N_x$ from $\mathcal{L}$. Note that doing this does not decrease $Ch$ for droplets, as that would entail scaling $\mathcal{L}$ proportionally with $N_x$ while weakening the forcing amplitude such that $Re_\lambda$ remains constant and $\eta$ is over-resolved (the approach of \cite{komrakova2015numerical}). We do not additionally pursue this as droplet dissolution is not significant in most of the parameter range considered in this study.

\begin{figure}
  \centerline{\includegraphics[width=0.85\linewidth]{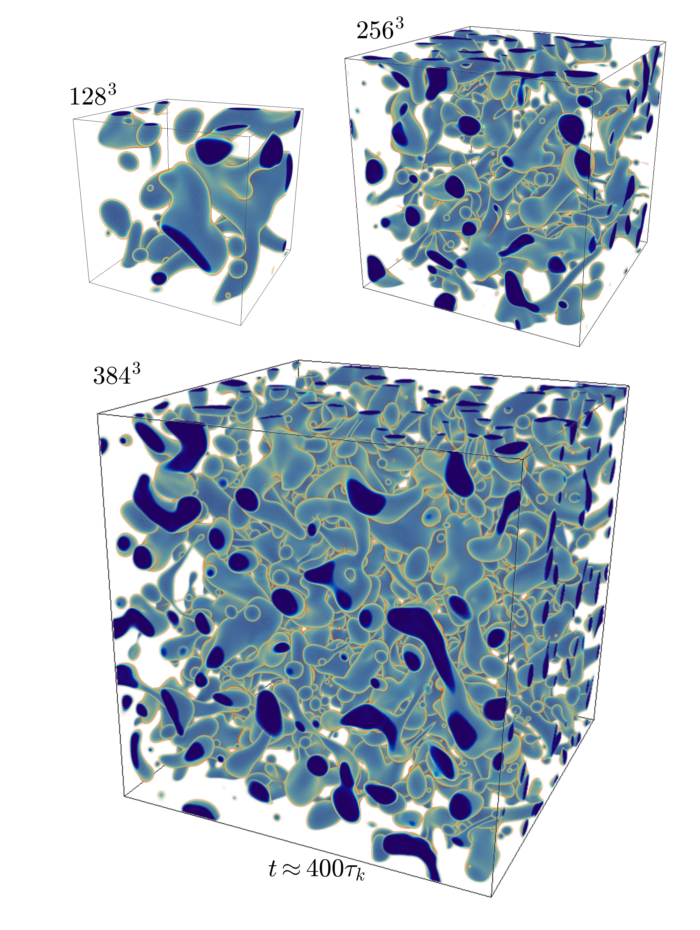}} 
  \caption{Volumetric droplet distribution for increasing domain sizes while maintaining the same energy density (power input) for cases D1, D2 and D3 with $N_x = 128^3, 256^3$ and $384^3$ respectively. The resolution of large droplet extensions becomes feasible at higher domain sizes. Here dark blue to orange goes from the droplet interior to the matrix phase.}
\label{fig:DomainSizeRenderIncreasingSize}
\end{figure}

Figure \ref{fig:DomainSizeRenderIncreasingSize} shows the droplets in the system (volume rendered) at $400\tau_k$ for increasing domain sizes. It can be seen that the largest structures in the $128^3$ domain span a significant fraction of the domain, whereas for increasing domain sizes the typical large scale structure becomes better resolved in relation to the domain size. The volume averaged droplet number density for these simulations was found to be almost identical.

\begin{figure}
  \centerline{\includegraphics[width=0.85\linewidth]{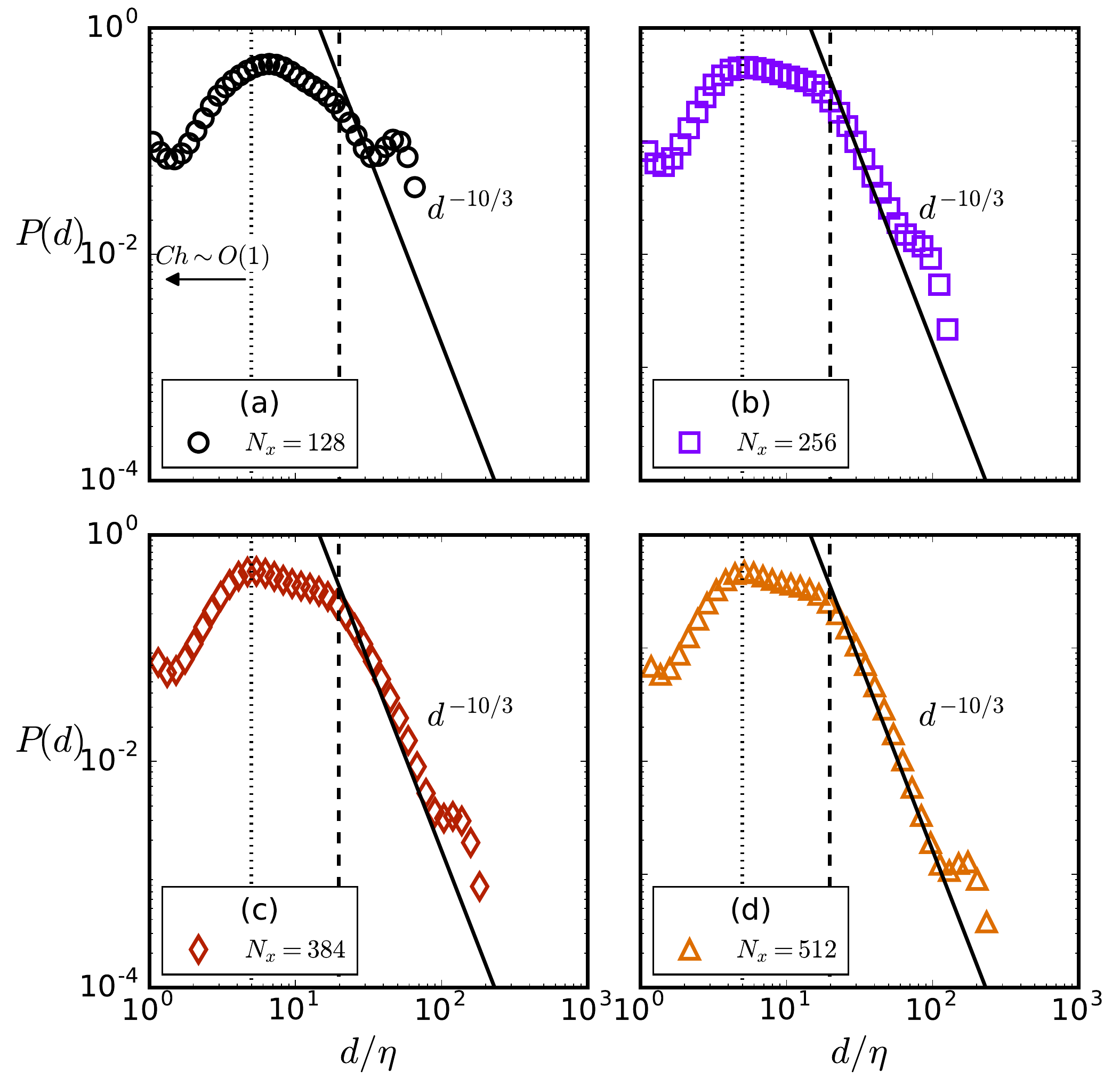}}
  \caption{Droplet size distributions for cases D1-D4, all with $\phi=0.15$ and $Re_\lambda\approx30$. The total number of droplets considered between times $150-600\ \tau_k$ are $\approx 5000, 40000, 54000, 133000$ for $N_x = 128, 256, 384, 512$ respectively. The dashed vertical line shows the Hinze scale and the dotted vertical line marks the limit $Ch\sim O(1)$.}
\label{fig:DomainSizeDropDistribution}
\end{figure}

The domain size limitation becomes apparent when considering the droplet distribution, as shown in figure \ref{fig:DomainSizeDropDistribution}. For the case of $N_x = 128^3$ (D1), the distribution is limited to a small region around the peak, clearly being cut off at a secondary peak emerging at higher $d/\eta$ due to a lack of resolution of larger structures. This case is under-resolved, the issue made acute with the small domain size, significant $\phi$ and moderate $Re_\lambda \approx 30$. We include this case to emphasize that the same issue might arise in simulations with higher $Re_\lambda$ and $N_x$ of high volume fraction dispersions. Upon increasing $N_x$, the distribution successively assumes a longer tail which closely follows the $d^{-10/3}$ scaling for droplets larger than the Hinze scale. 

\begin{figure}
	\begin{subfigure}{0.5\linewidth}
	\centerline{\includegraphics[width=\linewidth]{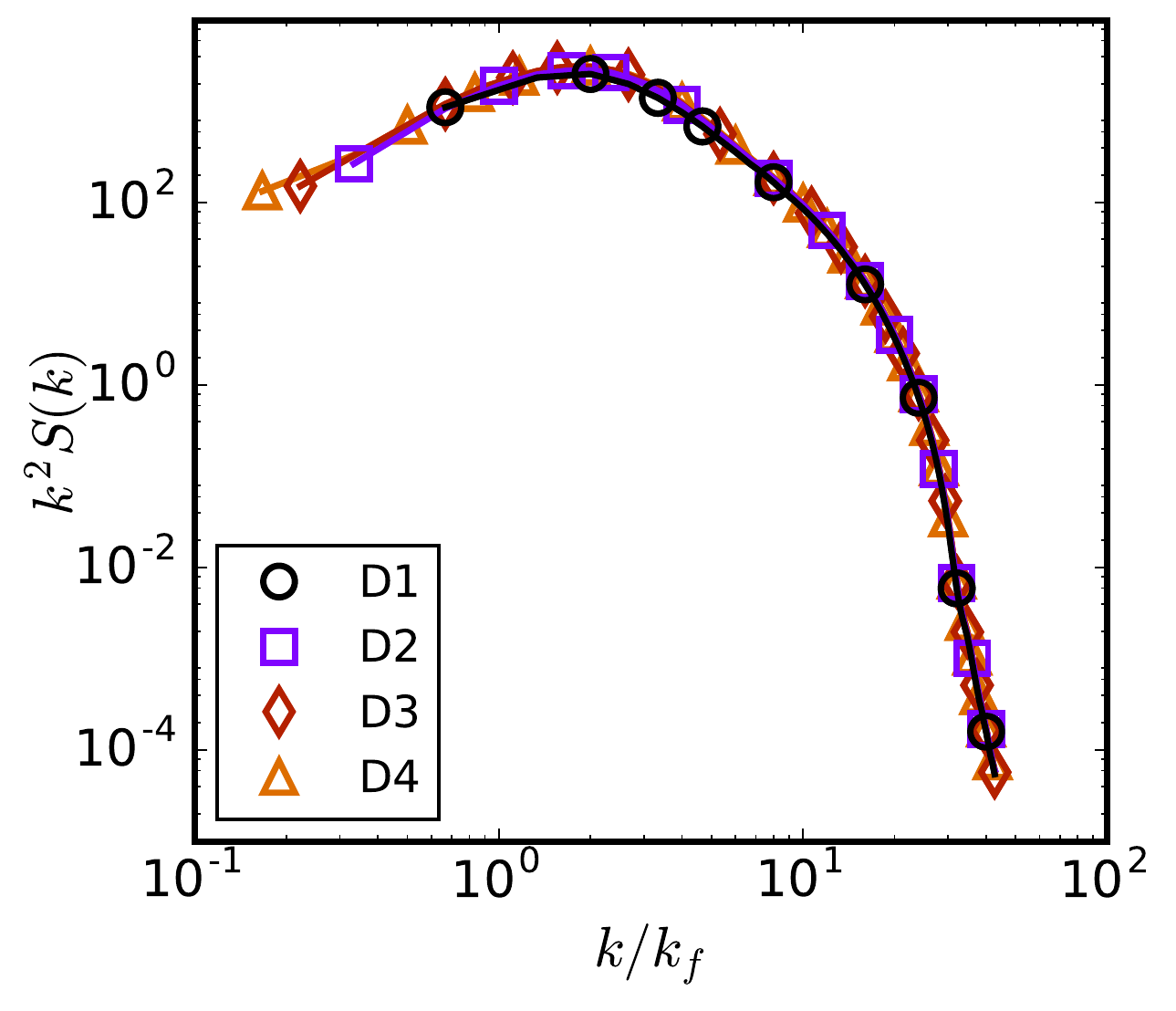}}
	\caption{Concentration spectrum}
	\label{fig:DomainSizeNxCompareStructureFactor}
	\end{subfigure} 
	\begin{subfigure}{0.5\linewidth}
  \centerline{\includegraphics[width=\linewidth]{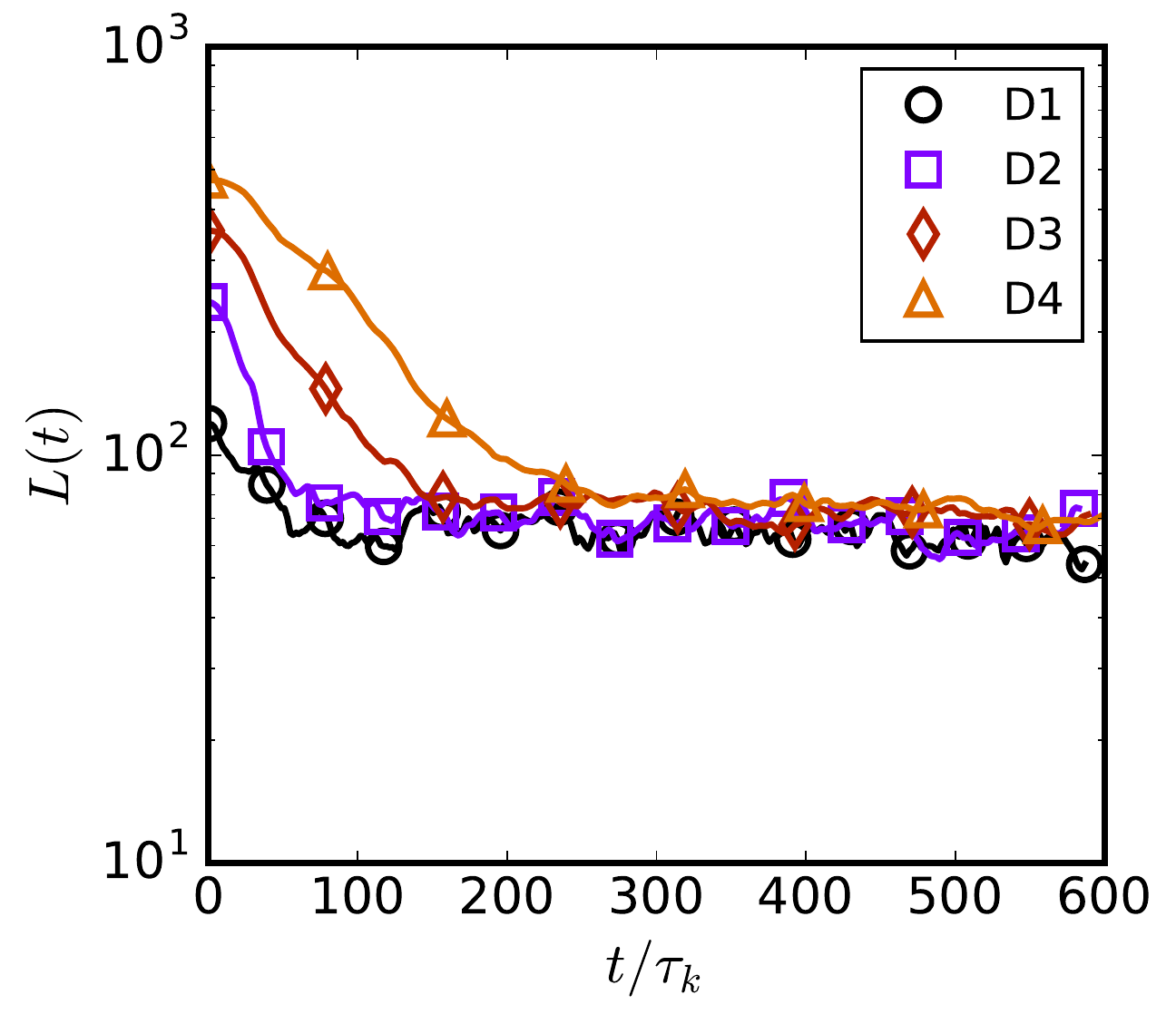}} 
  	\caption{Characteristic length evolution}
	\label{fig:DomainSizeNxCompareCharLen}
  \end{subfigure}
  \caption{Dispersion morphology characterized with the (a) concentration spectrum $k^2 S(k,t)$ and (b) characteristic length $L(t)$ for cases D1-D4. The concentration spectrum is averaged between times $150-600\tau_k$, sampled every $4\tau_k$. The importance of separation between the domain size $N_x$ or $\kmin$ and the dominant length scales characterizing the dispersion i.e. $k_d$ or $L(t)$ is evident from the fact that these two length scales can become comparable.}
\label{fig:DomainSizeNxCompare}
\end{figure}

Figure \ref{fig:DomainSizeNxCompareStructureFactor} shows the concentration spectrum for cases D1-D4, which first reflects the proper scaling as the spectra coincide for $k/k_f\geq 1$. The importance of resolving the dominant length scales characterizing the dispersion morphology vis-\`a-vis the domain size $N_x$ becomes apparent. The smallest wavenumber (largest length scale) that can be represented depends on $N_x$ as $\kmin = 2\pi/N_x$. For case D1, $\kmin$ is very close to the wavenumber corresponding to the peak in the concentration spectrum, i.e. the dominant wavenumber $k_d$ (or length scale $N_x/k_d$). If $\kmin \approx k_d$, two issues would tend to arise. First is that the dominant length scale of the emulsion morphology is comparable to the domain size making its dynamics under-resolved. Secondly, this structure will strongly interact with an image of itself due to periodicity of the domain, which is undesirable. For successively larger domains, the dominant length scale does not change (due to the same energy density across simulations). Further, the separation of $\kmin$ and $k_d$ is increased, which confirms that the largest structures ($\sim N_x/k_d$) are well resolved, while even larger structures (in the range of $k<k_d$) are formed but not sustained as the peak of $S(k)$ resides at $k_d$. The characteristic length evolution in figure \ref{fig:DomainSizeNxCompareCharLen} also shows that the morphology obtained for D1-D4 is similar, and that the typical length scale $L(t) \approx 80$ becomes better resolved in relation to the grid size upon increasing $N_x$.

\subsection{Effect of forcing wavenumber}\label{sec:forcingWavenumber}
\begin{figure}
  \centerline{\includegraphics[width=0.7\linewidth]{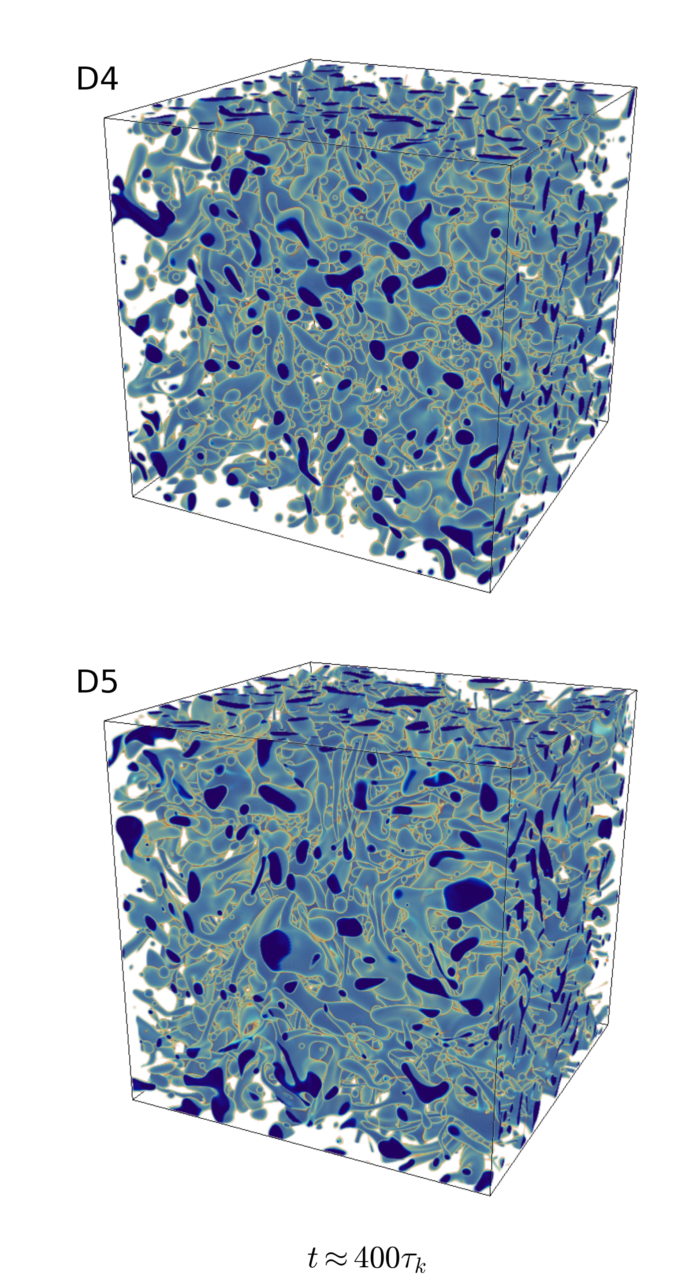}} 
  \caption{Volumetric droplet distribution for cases D4 and D5, where the forcing wavenumber is changed from $k_f=6$ to $k_f=1.5$, shown at $400\tau_k$. The D4 case shows a preponderance of smaller, more spherical droplets while D5 has more elongated filaments, possibly sustained due to the long wavelength of the forcing.}
\label{fig:DomainSizeRenderForcingWavenumberCompare}
\end{figure}

To highlight the consequences of forcing turbulence at the largest possible scale i.e. having $\mathcal{L}$ comparable to $N_x$ (hence maximizing $Re_\lambda$), we performed an additional simulation D5 with $k_f = 1.5\kmin$ and $\phi=0.2$ to compare with D4 ($k_f = 6\kmin$, $\phi=0.15$), while keeping the forcing amplitude the same, which results in $Re_\lambda = 118$ for case D5 (while $Re_\lambda=30$ for D4). Figure \ref{fig:DomainSizeRenderForcingWavenumberCompare} shows the typical morphology of the droplets (at a random time instance), where visibly the D4 case seems to have smaller, more spherical droplets, while D5 shows more elongated filaments. Despite the higher $Re_\lambda$, the dispersion does not comprise smaller droplets as droplet sizes depend on $\ang{\epsilon}$ which remains mostly unchanged. The presence of elongated filaments in D5 reflects the nature of the turbulence forcing. For a long cylindrical filament, a higher wavenumber forcing will generate more curvature variations. This would increase the possibility of filament breakup driven by Rayleigh-Plateau instabilities. A lower wavenumber forcing would generate weaker curvature differences in a long filament, and the timescale of breakup of these filaments might be comparable to the timescale of the large eddies, in which case the filaments will only break when the direction of the large scale shear changes. 

\begin{figure}
  \centerline{\includegraphics[width=\linewidth]{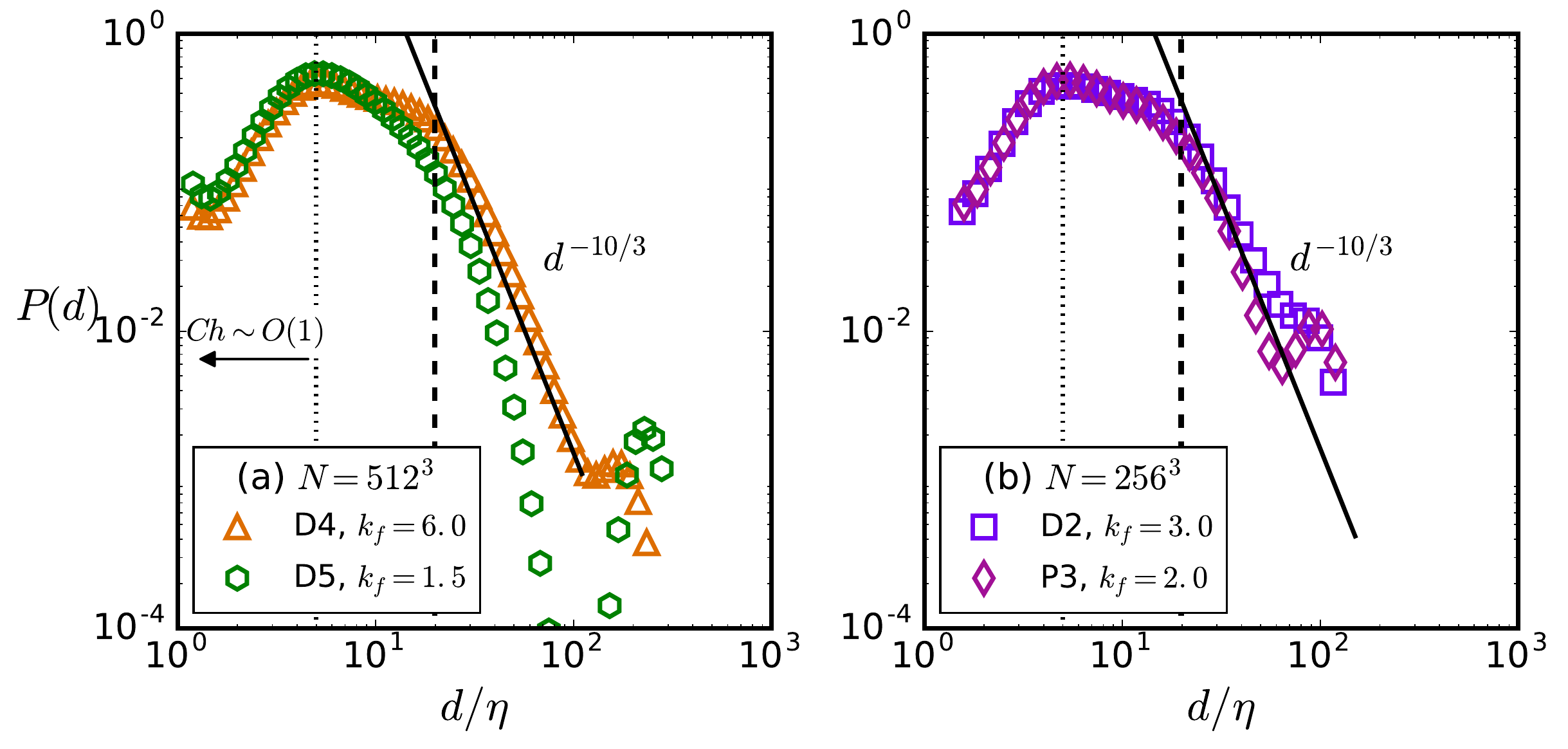}} 
  \caption{Droplet size distributions comparing (a) Cases D4 ($k_f = 6.0, \phi=0.15, Re_\lambda=30$) and D5 ($k_f = 1.5, \phi=0.20, Re_\lambda=118$), and (b) Cases D2 ($k_f = 3.0, Re_\lambda=30$) and P3 ($k_f=2.0, Re_\lambda=47$).}
\label{fig:DomainSizeDropDistributionForcingCompare}
\end{figure}

We further quantify the differences by calculating the droplet distribution for D4 and D5 (which have slightly different $\phi$), while also comparing simulations D2 (with $k_f=3.0$ and $Re_\lambda=30$) and P3 ($k_f=2.0$ and $Re_\lambda=47$) which have the same $\phi$, shown in figure \ref{fig:DomainSizeDropDistributionForcingCompare}. Indeed, the D5 case deviates from the $d^{-10/3}$ distribution above the Hinze scale reflecting the infrequent breakup of the long filaments that would lead to droplets in this range of sizes. This deficit of droplets shows up in a secondary peak at high $d/\eta$, which corresponds to the fewer, larger structures being sustained instead. A similar difference is seen between cases D2 and P3, where the P3 case shows a small peak at high $d/\eta$, again attributed to a lower wavenumber forcing. The same behaviour is reflected in the concentration spectrum as well between the cases (not shown here), where there is a relative increase in concentration at low wavenumbers for cases D5 and P3, although the characteristic length remains similar.

It is worthwhile to summarize the results from the domain size comparison and to draw conclusions. At modest $Re_\lambda$ ($<120$ in this study), the turbulence forcing wavelength and domain size influence the morphology. Having $N_x > \mathcal{L} \gg d$ (as in case D4) ensures sufficient resolution of the droplet breakup dynamics. While having $N_x \approx \mathcal{L} \gg d$ (case D5) causes the formation of longer filaments of the droplet fluid. Spatially, this causes the formation of larger droplets $d/\eta>100$ at the cost of some intermediate droplets $20<d/\eta<100$, for $d/\eta$ above the Hinze scale.

\section{Turbulent emulsion dynamics}\label{sec:dynamics}
\subsection{A quasi-equilibrium (limit) cycle}\label{sec:limitCycle}
Droplet number density plots such as figure \ref{fig:TurbVaryNumDropEvol} show oscillations of $N_d$ around a typical mean value which characterizes the dispersion morphology. So far, studies on droplets in turbulence refer to this state as a ``steady state'' where coalescence and breakup equilibrate. Since these oscillations can be significant (with its extreme values remaining bounded, similar to kinetic energy and dissipation), the dynamics should more accurately be called as a quasi-equilibrium (limit) cycle in the system state space comprising $\ang{E_k}$, $\ang{\omega^2}$, $\ang{E_\gamma}$ (i.e. the specific interfacial area multiplied with the interfacial tension $\gamma$) and $N_d$. Coalescence and breakup equilibrate in a statistical sense only, while the instantaneous dynamics is governed by temporal branches of alternating dominance of coalescence and breakup. Note that the term ``limit cycle'' is used loosely to illustrate the dynamics, since truly closed trajectories in phase space were not found, perhaps primarily due to intermittency and non-periodicity of the numerical solutions.

A dominant mediator of droplet breakup is intense enstrophy (or dissipation $\epsilon$). Since dissipation destroys turbulent kinetic energy, it is interesting to note that its interaction with the dispersed phase is associated with interfacial wrinkling, deformation and breakup - all mechanisms that increase the amount of surface energy in the system at the cost of kinetic energy. This excess energy, however, is still available in the flow field, and true destruction of it (i.e. into heat) must be mediated via kinetic energy dissipation, which occurs by the generation of smaller scales in the flow due to coalescence or damped oscillations of deformed droplet interfaces. A higher globally averaged $\ang{\omega^2}$ can be expected to increase the chance of droplet breakup (as it also reduces the effective Hinze scale), and vice-versa. Hence the trends seen in the $N_d$ evolution should reflect those in the evolution of $\ang{\omega^2}$, which in turn should follow the peaks and valleys of the kinetic energy $\ang{E_k}$ evolution.

\begin{figure}
  \centerline{\includegraphics[width=0.9\linewidth]{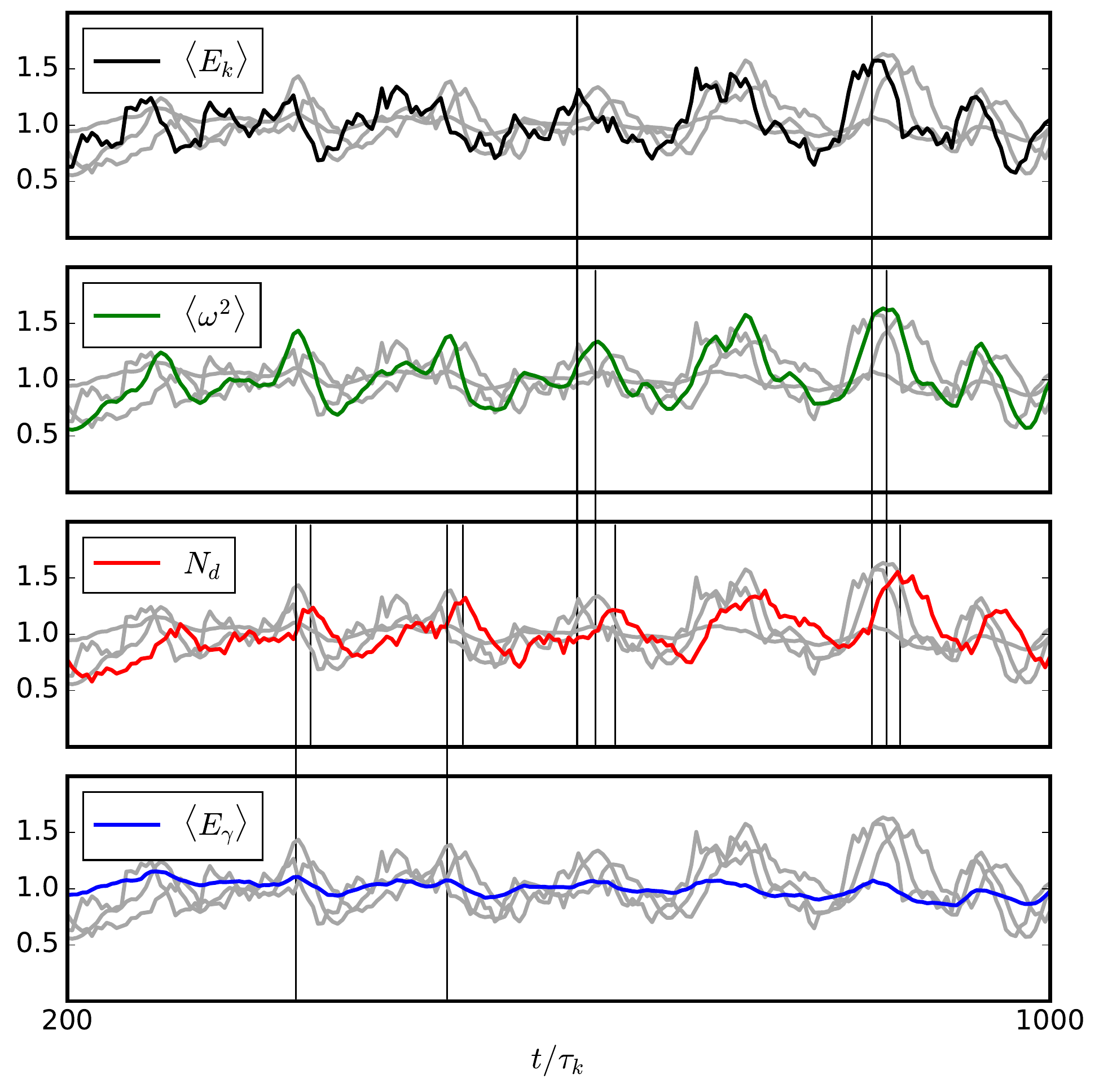}} 
  \caption{Evolution of quantities $N_d$, $\ang{E_k}$, $\ens$ and $\ang{E_\gamma}$ from case T5, where each quantity $\psi$ is normalized by its average value between $50\tau_k$ and $1000\tau_k$. The vertical lines spanning the top three columns mark typical instances where peaks in $\ang{E_k}$ lead to peaks in $\ens$ and subsequently $N_d$. Also, it is found that $\ang{E_\gamma}$ peaks prior to $N_d$, shown by the vertical lines spanning the bottom two panels.}
\label{fig:LimitCycleSignals}
\end{figure}

\begin{figure}
  \centerline{\includegraphics[width=\linewidth]{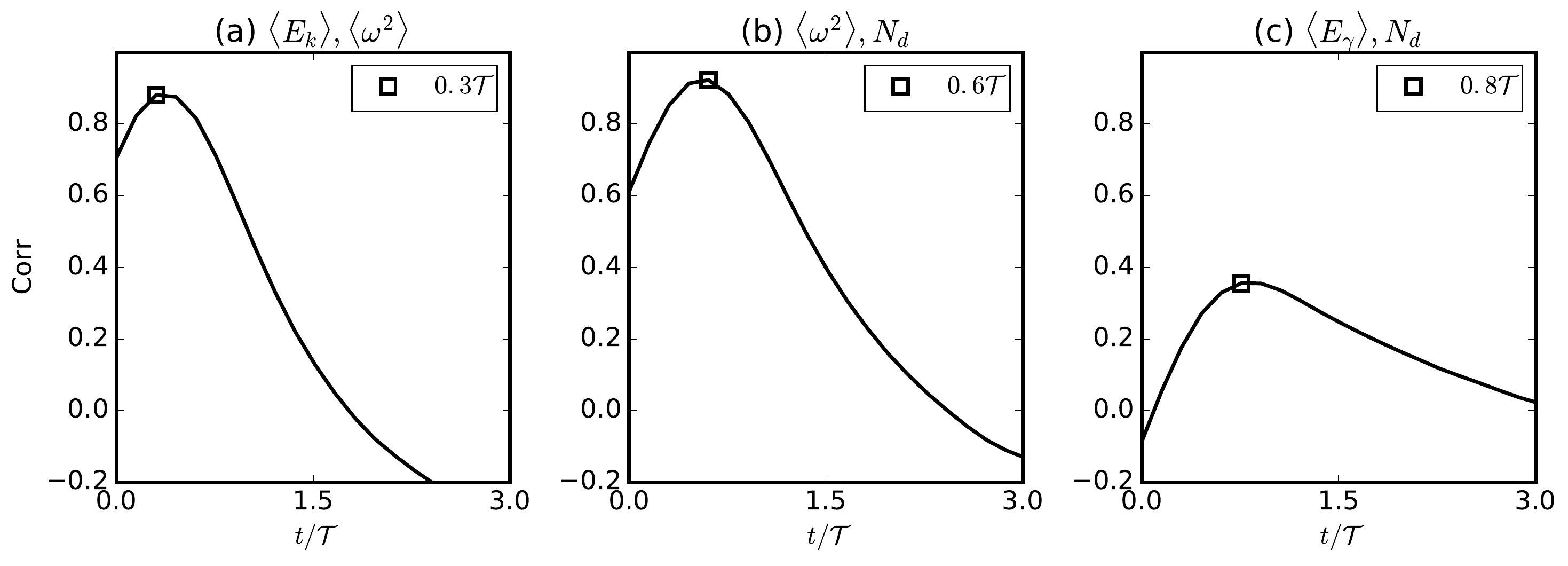}} 
  \caption{Correlation between $N_d$, $\ke$ and $\ens$ for case T5. $N_d$ consistently correlates strongly with $\ke$ with a temporal delay of $0.9\tstar$, while $\ang{E_\gamma}$ is found to attain its maximum value \textit{before} $N_d$, hinting that breakup occurs via extension of droplets into long filaments.}
\label{fig:LimitCycleCorrelations}
\end{figure}

This hypothesis is found to be true, and is shown in figure \ref{fig:LimitCycleSignals} as the evolution of $\ang{E_k}$, $\ang{\omega^2}$, $N_d$ and $\ang{E_\gamma}$ for case T5, where in the four panels each quantity has been separately highlighted in colour. Each variable has been further normalized by its time averaged value (between $50\tau_k$ and $1000\tau_k$) so that the evolution profiles become comparable. The vertical lines between the top three panels show two peaks of $\ang{E_k}$, which are reproduced in the $\ang{\omega^2}$ evolution and eventually in the $N_d$ evolution. A correlation can be calculated between the two signals as
\begin{equation}
\mathrm{Corr} = \frac{\overline{\ang{E_k(t)}\ang{\omega^2(t+\delta t)}}}{\overline{\ang{E_k}}\ \overline{\ang{\omega^2}}}
\end{equation}
where $\delta t$ is a time lag and the overbar is a temporal average. This has been done for the different signal pairs and is shown in figure \ref{fig:LimitCycleCorrelations}. Here $\ens$ is found to correlate strongly with $\ke$ with a time delay of $\sim 0.3\tstar$. $N_d$ shows a very strong correlation with $\ens$ at a time delay of $\sim 0.6\tstar$. Consequently, a significant correlation between $N_d$ and $\ke$ is found at $\sim 0.9\tstar$. The converse effect of droplets on turbulence can also be hinted at with this figure, where the valleys of the $N_d$ evolution invariably coincide with peaks in the $\ke$ evolution. This shows that when the droplet number density reduces due to coalescence, the excess surface energy is released into the flow as kinetic energy, which has been expounded by \cite{dodd2016interaction}. Since turbulence in our simulations is constantly forced (as opposed to \cite{dodd2016interaction} who simulate droplets in decaying turbulence)- the variation in $\ke$ in our simulations comes from a more complex confluence of the power input as well as the droplet dynamics.

We also observed this time delayed dynamics of $\ke$ and $N_d$ for cases with different parameters like turbulence forcing amplitude and interfacial tension, although for some cases the effect was less explicit. Particularly, for weaker $\gamma$ or lower $Re_\lambda$, the $N_d$ oscillations were not as extreme as for case T5 (where turbulence intensity and interfacial tension are both relatively stronger forces), although the $\ke$ and $N_d$ correlation was found to be strong. Generally, the dynamics can be described as follows. First the large scale structures generate higher velocity gradients at the dissipation scale (which may be due to the energy cascade if such exists) with an initial time lag. This larger dissipation rate is felt by the droplets, which respond by breaking up with a further time delay, increasing the number of droplets in the system. This process (from peaks in $\ke$ to peaks in $N_d$) was consistently found to take place with a delay of around $\sim 0.9\mathcal{T}$ across different cases, which is roughly the lifetime of the large eddies. This finding can be important for droplet dynamics models like population balance equations, where breakup kernels rely upon the instantaneous local value of $\epsilon$. If the temporal aspect to droplet populations is important, a relaxation time should separate cause and effect which is not done currently as seen in the various models reviewed by \cite{sajjadi2013review}. 

We also found that the surface energy $\ang{E_\gamma}$ peaks \textit{prior} to $N_d$ (last panel in figure \ref{fig:LimitCycleCorrelations}), which hints at the underlying breakup mechanism. Since generally daughter droplets together have a higher surface area than the parent droplet, $\es$ attaining its maximum values before $N_d$ suggests that droplets before breakup must form a rather elongated fluid filament, which has larger area than the daughter droplets formed after breakup.

\begin{figure}
  \centerline{\includegraphics[width=0.85\linewidth]{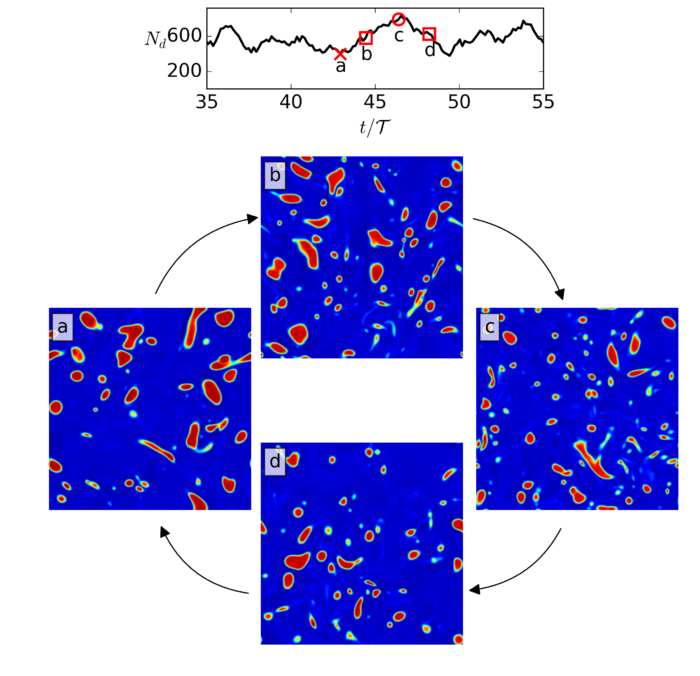}} 
  \caption{Quasi-periodic evolution of droplet morphology, cyclically visiting a typical state \textbf{`a'} marked by low $N_d$ (and hence low $\es$) and high $\ke$ and state \textbf{`c'} marked by large $N_d$ (and $\ang{E_\gamma}$) and low $\ke$. The transition from \textbf{`a'} to \textbf{`c'} happens via a dominance of breakup shown in state \textbf{`b'}, while the return from \textbf{`c'} to \textbf{`a'} via state \textbf{`d'} happens due to dominant coalescence. These snapshots are from case T5.}
\label{fig:LimitCycleMorphologies}
\end{figure}

In summary, the turbulent emulsion dynamics can also be interpreted as a quasi-periodic evolution in a state space comprising $\ke$, $\ang{\omega^2}$, $N_d$ and $\ang{E_\gamma}$. Essentially, there are two bounded extrema in the droplet number density at a given turbulent intensity for a certain set of fluid properties. These correspond to a state of low $N_d$ which is marked by fewer, relatively large droplets. When dissipation attains a subsequent peak, several of these droplets must be larger than the instantaneous Hinze scale - which leads to accelerated droplet breakup with takes the system to its other extremum - a state marked with high $N_d$. Most of the droplets in this state are stable and cannot undergo further breakup. As dissipation reduces, these droplets are advected around, and due to a higher chance of droplet-droplet collisions, coalescence dominates the next part of the state-space evolution. These two states also exhibit slightly different dispersion morphologies, as illustrated in figure \ref{fig:LimitCycleMorphologies}. The fluctuations in $N_d$ are caused by these two phases, where breakup and coalescence alternate in their dominance. In the $E_k-E_\gamma$ phase space, this can be viewed as (a somewhat erratic) evolution within a bounded region of finite $E_k$ and $E_\gamma$. We do find signatures of this behaviour, although to more accurately describe the $E_k-E_\gamma$ phase space requires further work where the contribution from breakup and coalescence are separately accounted for and the surface area is better resolved by simulating larger droplets in weaker turbulence. It should be noted, though, that the dynamics we report would correspond to local dynamics in larger droplet laden systems like stirred vessels or in clouds. When considering these systems as a whole, the equilibrium properties may not fluctuate as much as reported here, as the local fluctuations in different regions of the system would cancel out.

\subsection{Vorticity and interface alignment}\label{sec:vorticityAlign}
\begin{figure}
  \centerline{\includegraphics[width=\linewidth]{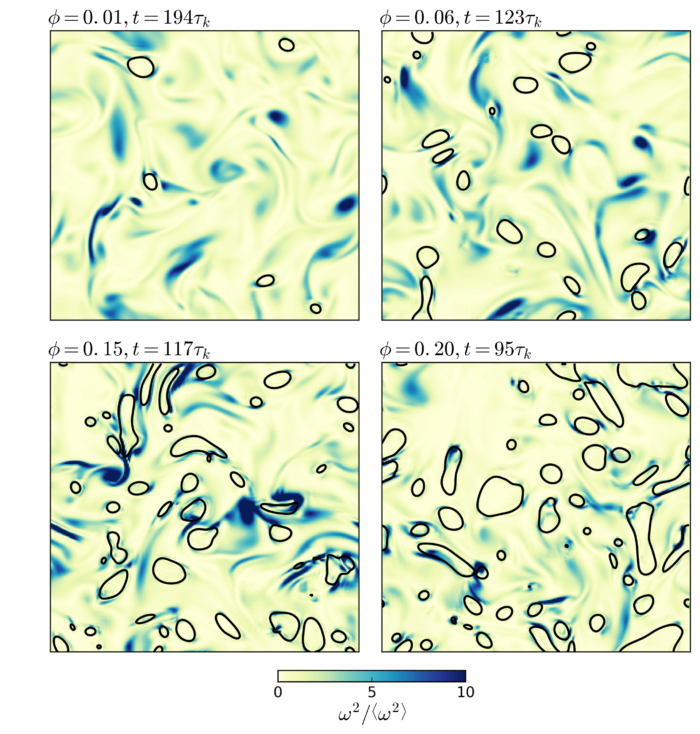}}
  \caption{Planar cross-sections (at $z=N_x/2$) of the enstrophy field $\omega^2$ normalized by the average enstrophy $\ang{\omega^2}$ along with droplet contours for varying $\phi$ values (cases P1-P4). These snapshots show the typical dissipation profiles with localized, intense dissipation events often concentrated around droplet interfaces or leading to droplet accretion.}
\label{fig:volFracVaryDissipation}
\end{figure}

Figure \ref{fig:volFracVaryDissipation} shows snapshots of enstrophy from a vertical cross-section of the varying $\phi$ simulations (P1-P4), with the droplet contours shown in black. Strong vortical regions are often found in the vicinity of the droplet interface and in the droplet wakes. There is strong interplay between the interfacial dynamics and dissipation, as strong vortical regions align with the interface \citep{shao2018direct} and cause wrinkling, and high local dissipative events can lead to droplet breakup \citep{perlekar2012droplet}.

\begin{figure}
  \centerline{\includegraphics[width=\linewidth]{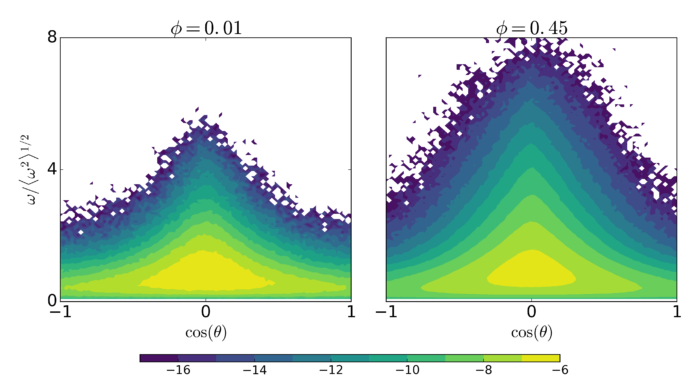}}
  \caption{Alignment between vorticity and the local interface normal is shown as the joint pdf of the cosine of the angle between them and the magnitude of vorticity, for the two extreme cases of $\phi=0.01,0.45$ (the other cases being qualitatively in between these two). The contour levels have been logarithmically spaced. Stronger vorticity tends to align orthogonal to the interface while weaker vorticity remains randomly aligned with the interface with a more uniform distribution.}
\label{fig:AlignmentVorticityInterfaceVolFrac}
\end{figure}

The angle between the vorticity vector and the interface normal can be quantified by using the distribution of the cosine of the angle between them. First, the density field $\rho_\beta$ is converted to a phase indicator field $\psi = (\rho_\beta - \rho_\beta^{\mathrm{out}})/(\rho_\beta^{\mathrm{in}} - \rho_\beta^{\mathrm{out}})$, such that $\psi=1$ in the droplet region, $\psi=0$ in the carrier fluid region, and $0<\psi<1$ at the interface. The typical phase indicator gradient then becomes $\boldsymbol{\nabla}\psi = 1/\dint$, and the cosine of the orientation angle is calculated where $\boldsymbol{\nabla}\psi > 0.01\dint$ (where $0.01$ ensures all the interfacial region is considered while ignoring the bulk regions where $\boldsymbol{\nabla}\psi=0$ by construction) as follows
\begin{equation}
\cos(\theta) = \frac{\boldsymbol{\nabla}\psi}{|\boldsymbol{\nabla}\psi|}\cdot\frac{ \boldsymbol{\omega}}{|\boldsymbol{\omega}|}
\end{equation}
where $\boldsymbol{\nabla}\psi/|\boldsymbol{\nabla}\psi|$ gives the interface normal. Recently, \cite{shao2018direct} showed using this measure that vorticity tends to align tangentially to droplet interfaces in turbulent flow. Here we extend their result in figure \ref{fig:AlignmentVorticityInterfaceVolFrac} which shows the joint probability distribution of the cosine of the orientation angle $\theta$ and the normalized vorticity vector $\boldsymbol{\omega}/\langle\omega^2\rangle^{1/2}$. Stronger vorticity is found to be more prone to align tangentially to the interface, which can be associated to a highly swirling motion in the orthogonal plane ($\cos (\theta) = 0$), which causes droplet accretion and subsequent tangential alignment of vorticity with interfaces. Weaker vorticity is incapable of exerting this influence on droplets, and hence would exhibit a uniform random distribution with respect to the interfaces. This result holds for droplets in the inertial range, while sub-Kolmogorov droplets might spin in local shear of the deep dissipation range.

\subsection{Effect of droplets on flow topology}\label{sec:topology}
\begin{figure}
  \centerline{\includegraphics[width=0.7\linewidth]{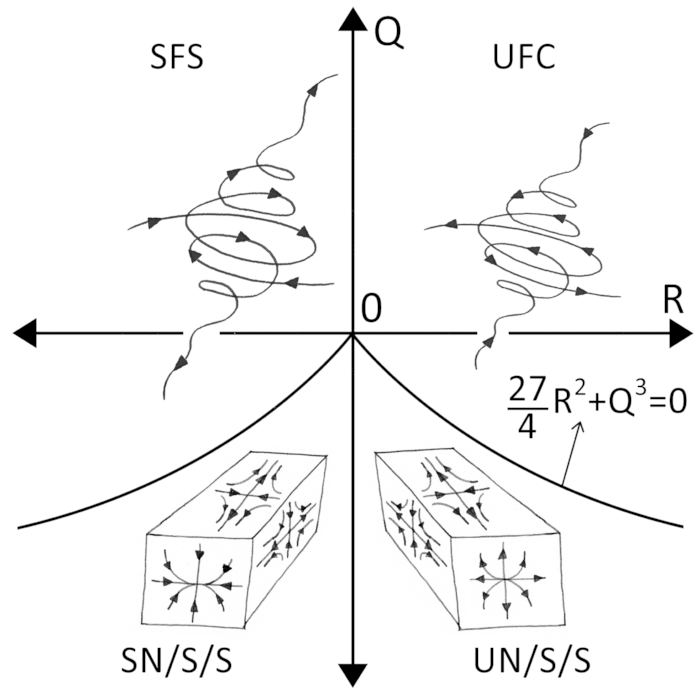}}
  \caption{The four distinct flow topologies of turbulent flow shown in the plane of $Q$ and $R$ i.e. the second and third invariants of the velocity gradient tensor $A_{ij}$. `SFS' is stable focus stretching, `UFC' is unstable focus compression, `SN/S/S' is stable-node/saddle/saddle and `UN/S/S' is unstable-node/saddle/saddle. This figure is an adaptation from the classification in \cite{ooi1999study}.}
\label{fig:qrjpdf-schematic}
\end{figure}

Local flow topology is described in terms of the three invariants ($P$, $Q$ and $R$) of the velocity gradient tensor $A_{ij}=\partial u_i/\partial x_j$, which form the coefficients of its characteristic equation
\begin{equation}
\lambda^3 + P\lambda^2 + Q\lambda + R = 0
\end{equation}
where $P = -A_{ii}$, $Q = -A_{ij}A_{ji}/2$ and $R = -A_{ij}A_{jk}A_{ki}/3$. For incompressible flow, $P=0$ (i.e. the sum of the eigenvalues). In the $P=0$ plane (or the $QR$-plane), turbulent flow of diverse kinds produces a teardrop-like profile for the joint probability distribution of $Q$ and $R$ with four distinct flow topologies that have been illustrated in figure \ref{fig:qrjpdf-schematic} (adapted from \cite{ooi1999study}). The curve $D = 27R^2/4 + Q^3 = 0$ divides the region with three real eigenvalues of $A_{ij}$ (below, where $D<0$) from the region with one real and a pair of complex conjugate eigenvalues (above, where $D>0$). The most dominant flow features are stable focus stretching `SFS' (i.e. vortex stretching) and unstable-node/saddle/saddle `UN/S/S' i.e. bi-axial straining \citep{chacin2000dynamics}. `UFC' corresponds to unstable focus compression (or vortex compression) and `SN/S/S' is stable-node/saddle/saddle (or axial straining).

The presence of droplets or particles which interact with the flow can modify the distribution of flow topologies, which is a modification of turbulence structure at a more local and fundamental level than for instance modifications to the kinetic energy spectrum. This has been well investigated for particle laden turbulence \citep{rouson2001preferential,bijlard2010direct} and recently shown for elastic polymers in turbulence by \cite{perlekar2010direct}. The effect of the latter can be similar to droplets which themselves are elastic objects due to interfacial tension, with the additional complexity of breakup and coalescence. Recently, \cite{shao2018direct} showed a mild suppression of bi-axial straining in droplet laden turbulence upon changing the Weber number.

\begin{figure}
  \centerline{\includegraphics[width=0.9\linewidth]{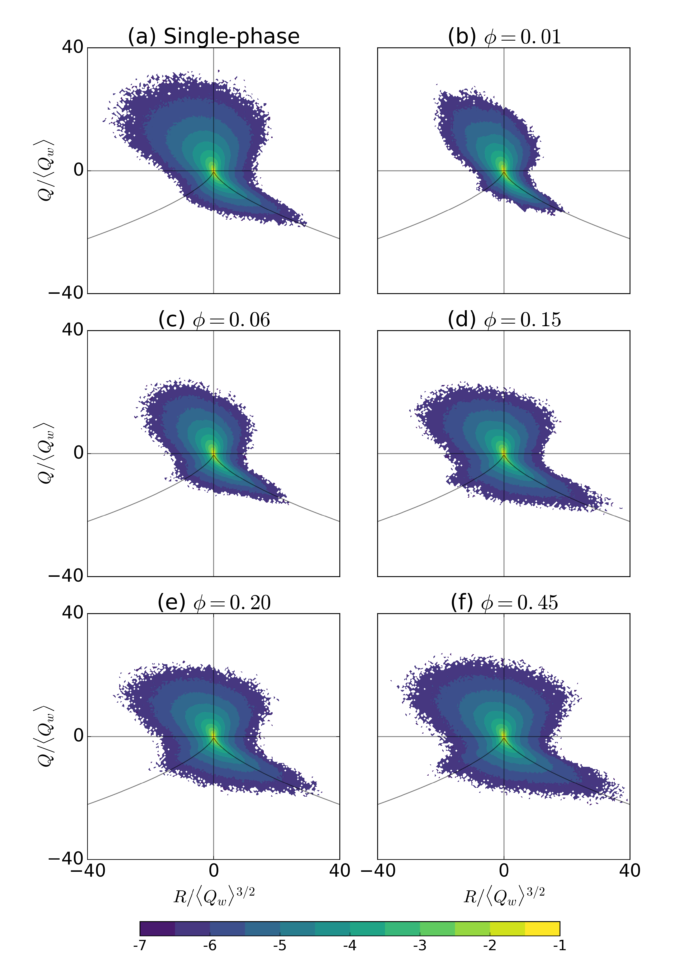}}
  \caption{Joint PDFs of the second and third invariants ($Q$ and $R$) of the velocity gradient tensor shows the typical teardrop profile characteristic of single-phase turbulence being modified into a more symmetric profile with an increase in axial straining and vortex compression. Here $\langle Q_w \rangle = \langle \omega^2 \rangle/4$ and the quantities are calculated over the entire multiphase velocity field, sampled at $5$ time instances separated by $20\tau_k$. The solid lines mark $Q=0$, $R=0$ and $D= 27R^2/4 + Q^3 = 0$, and the contour levels have been logarithmically spaced.}
\label{fig:qrjpdf-PhiVary}
\end{figure}

How droplets modify flow topology has not fully been investigated so far. Here, we first show the influence of increasing dispersed phase volume fraction on the $QR$ profiles calculated using simulations P1-P5. Since $Re_\lambda$ for these cases varies (and is almost a factor 2 lower than the corresponding single-phase turbulence simulation, see table \ref{tab:TurbEmPhiVary}), the normalization factor $\langle Q_w\rangle = \langle \omega^2 \rangle/4$ \citep{ooi1999study} is calculated for each case separately. This allows us to focus on the modification of flow features alone, without comparing the magnitude of these extreme $QR$ events. Figure \ref{fig:qrjpdf-PhiVary} shows the $QR$ field sampled over the entire multiphase velocity field. For case (b) $\phi=0.01$, the profile is narrower than for single-phase turbulence, case (a), although the overall shape is similar. This might be due to the $\phi=0.01$ dispersion being dilute, which makes coalescence infrequent. Overall, in this case, the flow field is similar to that in single-phase turbulence, and coalescence generated smaller scale features are rare. This seems likely, as at successively higher volume fractions, cases (c) through (f), the $QR$ profile is influenced more significantly and it tends to become more symmetric across the $R=0$ line. This follows from an increase in the axial straining part of the flow, along with an extension of the profile into the $D>0$ and $R>0$ region which shows a relative increase in vortex compression as opposed to vortex stretching ($D>0$ and $R<0$). 

\begin{figure}
  \centerline{\includegraphics[width=0.9\linewidth]{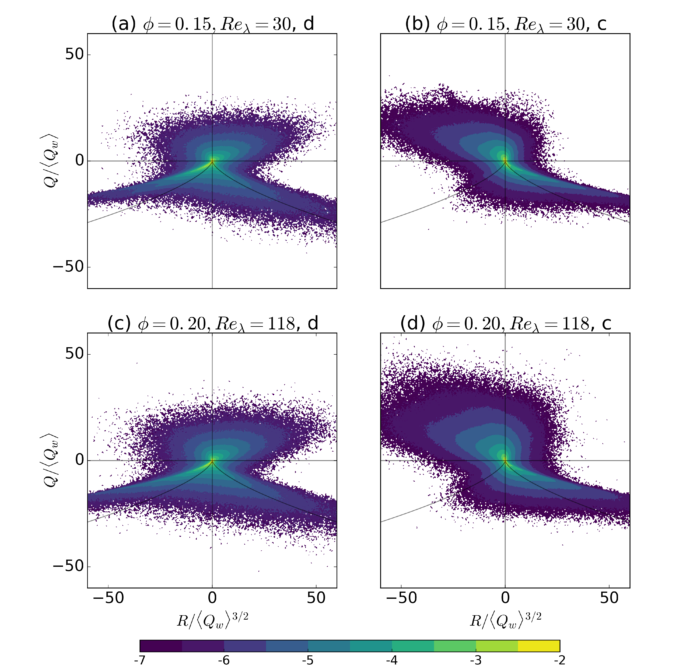}}
  \caption{Joint PDFs of $QR$ sampled in the droplet phase (``d'') and continuous phase (``c'') for cases D4 and D5. The $QR$ profile appears to flip on the $R=0$ axis for the droplet phase, with a striking increase in vortex compression and axial straining. The continuous phase $QR$ profile remains mostly tear-drop like with minor increase in axial straining.}
\label{fig:qrjpdf-inoutcompare}
\end{figure}

Modification of the $QR$ profile due to an increase in $\phi$ hints that it is a consequence of turbulence being constrained by the dispersed phase. To validate this claim, in figure \ref{fig:qrjpdf-inoutcompare} the $QR$ profiles are shown while being sampled inside and outside the droplet regions (marked as ``d'' for droplet-phase and ``c'' for continuous-phase). This has been done for simulations D4 and D5 (which have the highest resolution, and significantly different $Re_\lambda=30$ and $118$ respectively). The $QR$ profiles have been sampled at 5 time instances separated by $100\tau_k$. The difference between the flow topology in the droplet and continuous phase is striking, where within the droplet region $QR$ profile seems to almost have flipped across the $R=0$ axis. There is a significant increase in axial straining and vortex compression inside the droplets. This may be ascribed to the presence of interfaces surrounding droplets which behave like elastic surfaces. Vortices being stretched inside the droplets will try to elongate the droplet along the stretching axis, and this will be counteracted by interfacial tension which would instead tend to compress vortices. Since vortex compression contributes to energy dissipation \citep{tsinober2009informal}, an enhancement of energy dissipation might be expected inside droplets from these results. Further investigation of this is left for future work. The continuous phase $QR$ profile remains mostly tear-drop like, with minor increase in axial straining and vortex compression.  

\section{Conclusions}
We perform direct numerical simulations of emulsions under homogeneous, isotropic turbulence conditions performed by using the pseudopotential lattice-Boltzmann method. New findings on droplet size distributions, multiphase kinetic energy spectra, coupled kinetic energy and droplet number density dynamics, interface-dissipation interactions and modification of turbulence flow topology in emulsions are reported. 

The process of dispersion formation is investigated for varying volume fractions of the dispersed phase and varying turbulence intensities for an emulsion with a density and viscosity ratio of 1. Using an appropriate set of parameters (such that the pseudopotential repulsive force between components dominates the local turbulence force), the effect of droplet dissolution is mitigated, an issue that was found limiting in previous work \citep{perlekar2012droplet,komrakova2015numerical}. While further maintaining spurious currents to well below the physical velocity scales, the multiphase kinetic energy spectra were shown to exhibit signatures of breakup and coalescence at wavenumbers smaller and larger than the inverse Hinze scale respectively. 

At small wavenumbers, energy is primarily extracted from the flow, where a higher dispersed phase volume fraction $\phi$ extracts more energy due to the profusion of interfaces. At large wavenumbers, for successively higher $\phi$, the energy content of the dissipation range increases due to more frequent coalescence which generates smaller scale motions. The droplet distribution is shown to follow the \cite{deane2002scale} $d^{-10/3}$ scaling above the \cite{hinze1955fundamentals} scale.

The importance of the relative resolution between the various length scales that govern turbulence droplet simulations is emphasized. We show that it is important to resolve $N_x>\mathcal{L}$ to correctly capture droplet deformation and breakup at relatively weaker turbulence intensities and high volume fractions, where otherwise the droplet fluid can form a complex tangle of elongated filaments as the maximum droplet deformation becomes unresolved. We also maintain that $\mathcal{L}\gg d \gg \eta$, such that the droplets interact mainly with the inertial range of turbulence.

In line with recent results \citep{shao2018direct}, vorticity is shown to strongly align tangentially to droplet interfaces. This effect was shown to be stronger for higher vorticity magnitudes. The presence of dispersed phase is also shown to significantly alter the flow topology represented by the joint pdf of $QR$, i.e. the second and third invariants of the velocity gradient tensor, much more acutely than recognized \citep{shao2018direct}. The well known tear-drop like profile becomes almost flipped across the $R=0$ axis when sampled inside the droplet in comparison to sampling in the carrier phase. A striking increase in axial straining and vortex compression is found in the droplets, which hints at an interplay of interfacial tension which tries to counteract any extensional vortical motions. This result hints that droplets might cause enhanced dissipation in their interior. The carrier fluid topology retains features of the well known tear-drop profile \citep{chacin2000dynamics} with only minor increase in axial straining and vortex compression.

Last but not the least, we show for the first time the dynamics of the quasi-equilibrium between coalescence and breakup under constant energy input to the system which leads to sustained turbulence over very long simulation times (around $100\mathcal{T}$). This state is often called a ``steady state'', although the dynamics more closely resembles a limit-cycle in the state-space of kinetic energy $\ang{E_k}$, enstrophy $\ang{\omega^2}$, droplet number density $N_d$ and surface energy $\ang{E_\gamma}$. The extreme values of $\ang{E_k}$ manifest in the $\ang{\omega^2}$ evolution with a certain time delay, which then again show up in the $N_d$ evolution leading to a time-delayed dynamics. The dispersion oscillates between two morphologies, the journey between them being mediated by alternating bouts of dominant breakup and coalescence. Surface energy was found to peak prior to droplet breakup, reflecting the underlying breakup mechanism which involves the stretching of droplet fluid filaments, which have a higher surface area than the subsequently formed daughter droplets. 

We believe that this time delayed dynamics will be found in localized regions of much larger droplet laden systems, where the overall system may not exhibit significant fluctuations in state-space variables, as the localized fluctuations would cancel each other. However, in smaller, finite systems (as prevalent in turbulence resolving droplet laden simulations \citep{elghobashi2019dns}), this can be an important consideration, as the ``steady state'' can have its own interesting dynamics. These considerations of delayed temporal dynamics may also be relevant to developing more realistic breakup and coalescence kernels which currently correlate state-space variables instantaneously \citep{sajjadi2013review}, which we have not explored given the limits of the current work.

Further investigation of the system evolution in the $\ang{E_k}-\es$ phase space would help describe the exact exchange of energy, where the effects of coalescence and breakup would need to be isolated. This may be done by simulating larger droplets in weak turbulence, which would correspond to a detailed view on individual droplets near the dissipation range, and it is something we wish to investigate in the future. 

We hope that this paper brings to attention the avenue of considering the details of resolved simulations from different perspectives (as we have attempted, while considering the limitations of our work). This helps reinforce our understanding of the phenomena at different levels. A statistical perspective (looking at spectra, time averaged quantities etc) helps with an overall description, while a dynamical systems perspective on the state-space helps pave the way for deciphering the true mediation of cause and effect like droplet-dissipation interactions and the modification of turbulence due to droplets, which we are only beginning to now understand.

\section*{Acknowledgements}
SM would like to thank Jason Picardo (International Center for Theoretical Sciences, Bangalore), Luis Portela (TU Delft) and Jos Derksen (University of Aberdeen) for insightful discussions regarding this work. This work is funded by the Institute for Sustainable Process Technology (ISPT), the Netherlands. 

\appendix
\section{}\label{app:Threshold}
In this section we briefly discuss the segmentation of droplets. The simulations output a continuous density field for both components $\alpha$ and $\beta$. As mentioned, the density variation of a component indicates the presence of droplets, where the density of component $\beta$ inside the droplet $\rho_\beta^\mathrm{in} \approx 4.4$ and that outside the droplet $\rho_\beta^{\mathrm{out}} \approx 0.4$ [lu] during the simulations. The droplet identification is done by picking a threshold density value $\rho^{\mathrm{c}}$, and every contiguous region with $\rho_\beta > \rho^{\mathrm{c}}$ is identified as a droplet, using a spatial segmentation (or clustering) algorithm previously developed by \cite{siebesma2000anomalous}. The results, hence, should be independent of $\rho^{\mathrm{c}}$.

\begin{figure}
  \centerline{\includegraphics[width=0.75\linewidth]{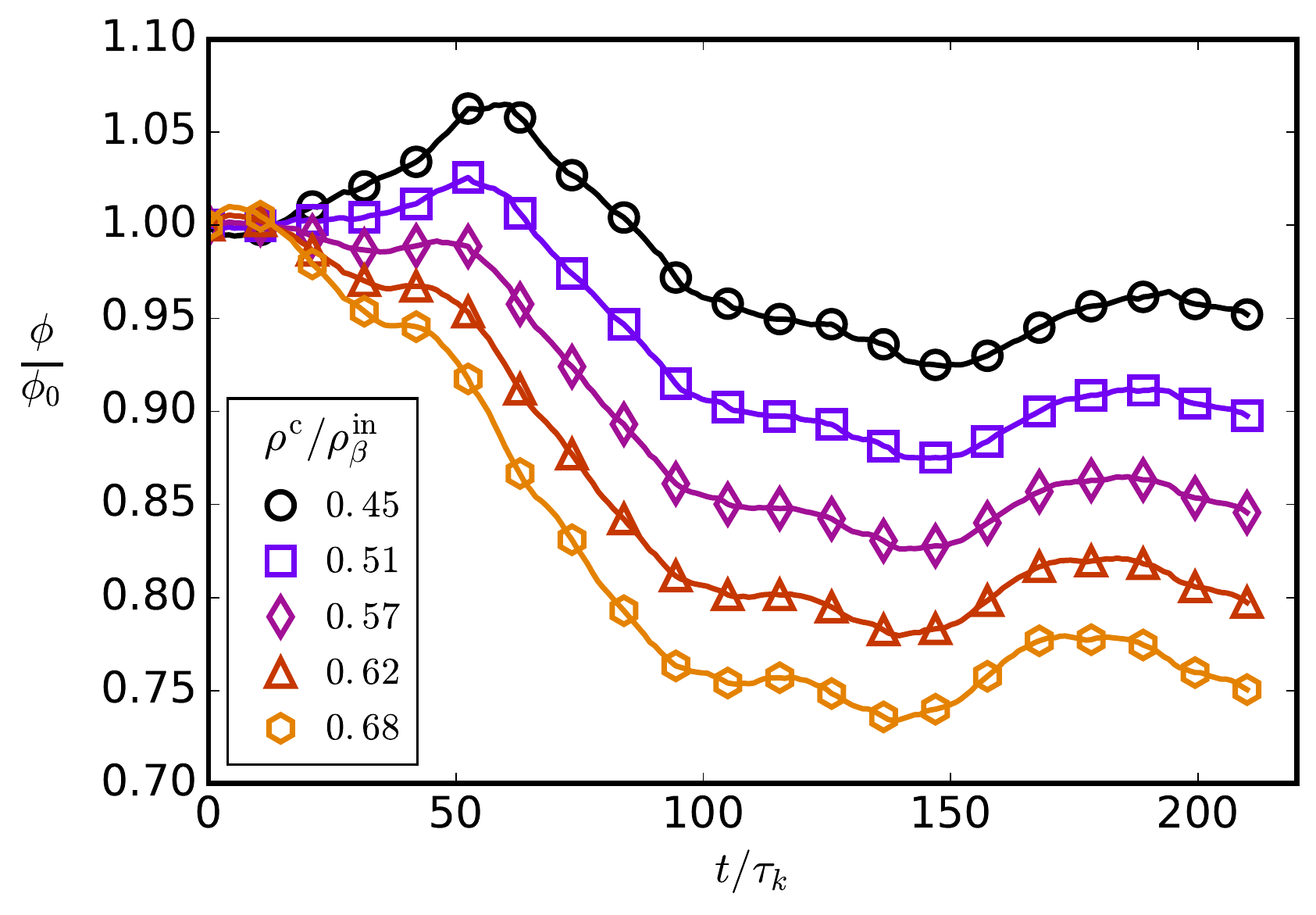}} 
  \caption{Normalized phase fraction evolution for varying $\rho^{\mathrm{c}}$ used to segment droplets. Here $\rho^{\mathrm{c}}$ is reported as a fraction of $\rho^{\mathrm{in}}_\beta$, while $\rho^{\mathrm{out}}_\beta$/$\rho^{\mathrm{in}}_\beta = 0.1$, therefore the useful range of $\rho^\mathrm{c}/\rho^{\mathrm{in}}_\beta$ is $0.1$ to $1.0$, and $\rho^{\mathrm{c}}/\rho^{\mathrm{in}}_\beta=0.55$ is halfway.}
\label{fig:threshVolFrac}
\end{figure}

Figure \ref{fig:threshVolFrac} shows the relative evolution of the volume fraction over time for the case $\phi=0.06$ for different threshold values shown as a fraction of $\rho^{\mathrm{in}}_\beta$, while $\rho^{\mathrm{out}}_\beta/\rho^{\mathrm{in}}_\beta = 0.1$. This makes the useful range of thresholding values $\rho^\mathrm{c}/\rho^{\mathrm{in}}_\beta \in \left[0.1, 1.0\right]$, where $\rho^{\mathrm{c}}/\rho^{\mathrm{in}}_\beta=0.55$ is halfway. Lower values of $\rho^\mathrm{c}$ create for slightly larger droplet regions, whereby $\phi/\phi_0$ increases. The real loss in $\phi$ is due to the dissolution of small droplets, and partially due to a loss in the droplet phase density to compensate for the increase in overall droplet surface area after breakup. This is affirmed by the increase in $\phi$ upon lowering $\rho^\mathrm{c}$ so that more of the interfacial region is considered to lie inside the droplets.

\begin{figure}
  \centerline{\includegraphics[width=\linewidth]{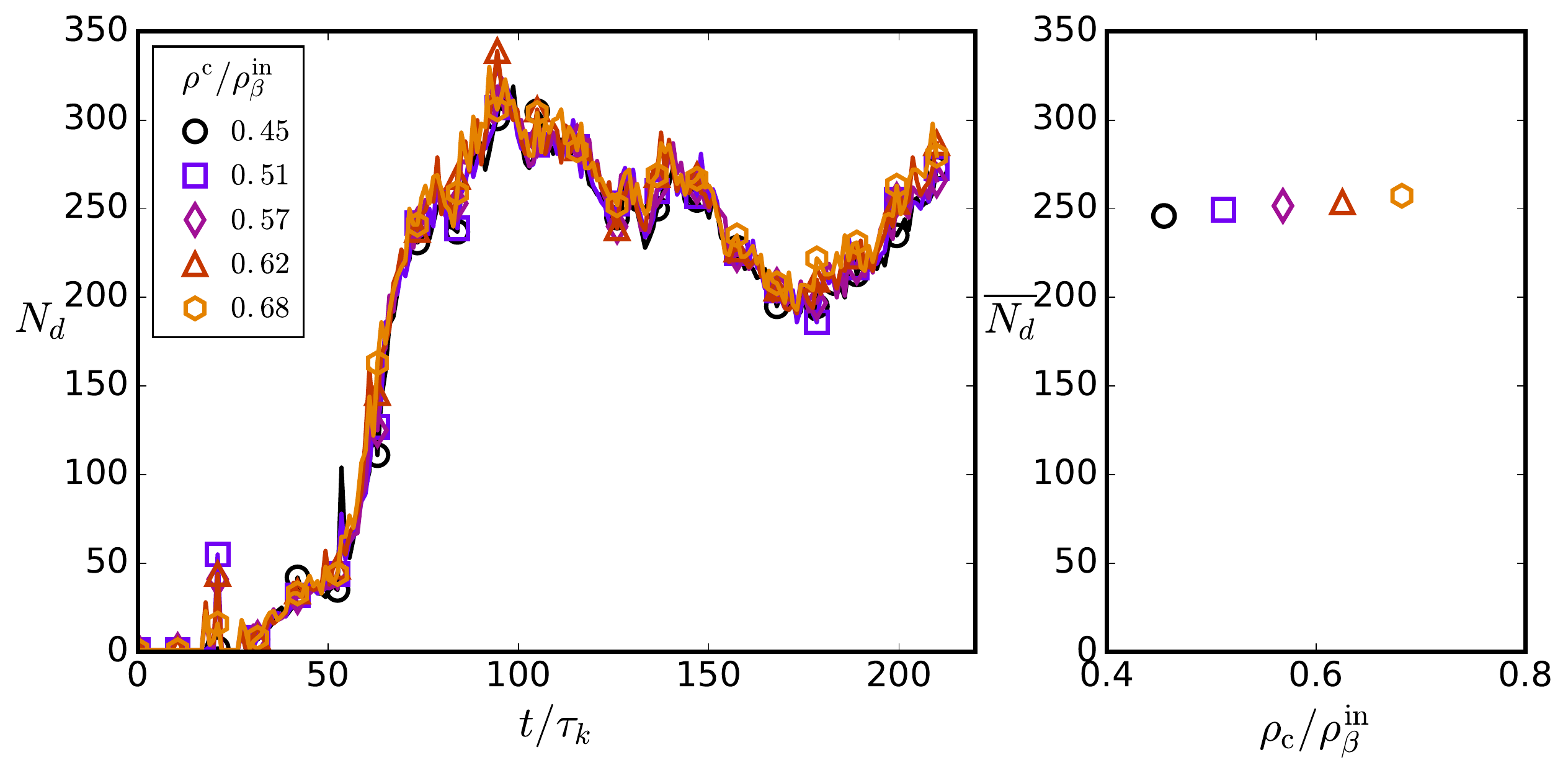}} 
  \caption{Evolution of droplet number density $N_d$ for varying $\rho^{\mathrm{c}}$ shows that the number of droplets identified is almost independent of $\rho^{\mathrm{c}}$.}
\label{fig:threshNdEvolution}
\end{figure}

Figure \ref{fig:threshNdEvolution} shows the evolution of the number of droplets $N_d$ in the system for different threshold magnitudes, which is seen to have minimal influence on $N_d$. Similarly, the droplet distribution was also found to be virtually unaffected by the choice of $\rho^{\mathrm{c}}$ as long as it lies within the droplet interface. The results reported in this work use $\rho^{\mathrm{c}}/\rho^{\mathrm{in}}_\beta=0.57$, which is very close to the halfway value.

\bibliographystyle{jfm}
\bibliography{turbEm}

\end{document}